\documentclass[aps,reprint,showpacs]{revtex4-1}

\usepackage{graphicx}
\usepackage{upgreek}

\usepackage{amssymb}
\usepackage{amsmath}
\usepackage{bbold}

\renewcommand{\vec}[1]{{\mathbf #1}}

\begin{document}


\title{Finite-size scaling analysis of localization transition\\
for scalar waves in a 3D ensemble of resonant point scatterers}



\author{S.E. Skipetrov}
\affiliation{Universit\'{e} Grenoble Alpes, LPMMC, F-38000 Grenoble, France}
\affiliation{CNRS, LPMMC, F-38000 Grenoble, France}
\date{\today}

\begin{abstract}
We use the random Green's matrix model to study the scaling properties of the localization transition for scalar waves in a three-dimensional (3D) ensemble of resonant point scatterers. We show that the probability density $p(g)$ of normalized decay rates of quasi-modes $g$ is very broad at the transition and in the localized regime and that it does not obey a single-parameter scaling law for finite system sizes that we can access. The single-parameter scaling law holds, however, for the small-$g$ part of $p(g)$ which we exploit to estimate the critical exponent $\nu$ of the localization transition. Finite-size scaling analysis of small-$q$ percentiles $g_q$ of $p(g)$ yields an estimate $\nu \simeq 1.55 \pm 0.07$. This value is consistent with previous results for Anderson transition in the 3D orthogonal universality class and suggests that the localization transition under study belongs to the same class.
\end{abstract}

\maketitle

\section{Introduction}
\label{intro}

Anderson transition is a transition in transport properties of a disordered quantum or classical wave system \cite{anderson58,evers08,lagendijk09,abrahams10}. It is due to destructive interferences of scattered waves leading to formation of spatially localized eigenstates and halt of wave transport through the system. In the most common case of time-reversal symmetric and spin-rotation invariant systems, the transition exists in three dimensions (3D) whereas all states are localized in lower dimensions for arbitrary weak disorder \cite{abrahams79}. Experimental evidences for Anderson transition in 3D were found in the low-temperature electrical conductance of disordered solids \cite{rosenbaum80, paalanen82}, transmission \cite{hu08} and reflection \cite{cobus15} of elastic waves from disordered media, phase-space dynamics of cold atoms in a quasi-periodic force field \cite{chabe09}, and real-space expansion of ultra-cold atomic clouds in optical speckle potentials \cite{jendr12}. The search for Anderson transition in optical systems has not been conclusive despite a considerable experimental effort during the last two decades \cite{wiersma97,vanderbeek12,storzer06,sperling13,sperling16,skip16}. Localization transition may be difficult or even impossible to reach for light due to near-field effects in dense disordered media required to achieve strong scattering \cite{skip14}.

\begin{figure}
\vspace{-4mm}
\includegraphics[width=0.97\columnwidth]{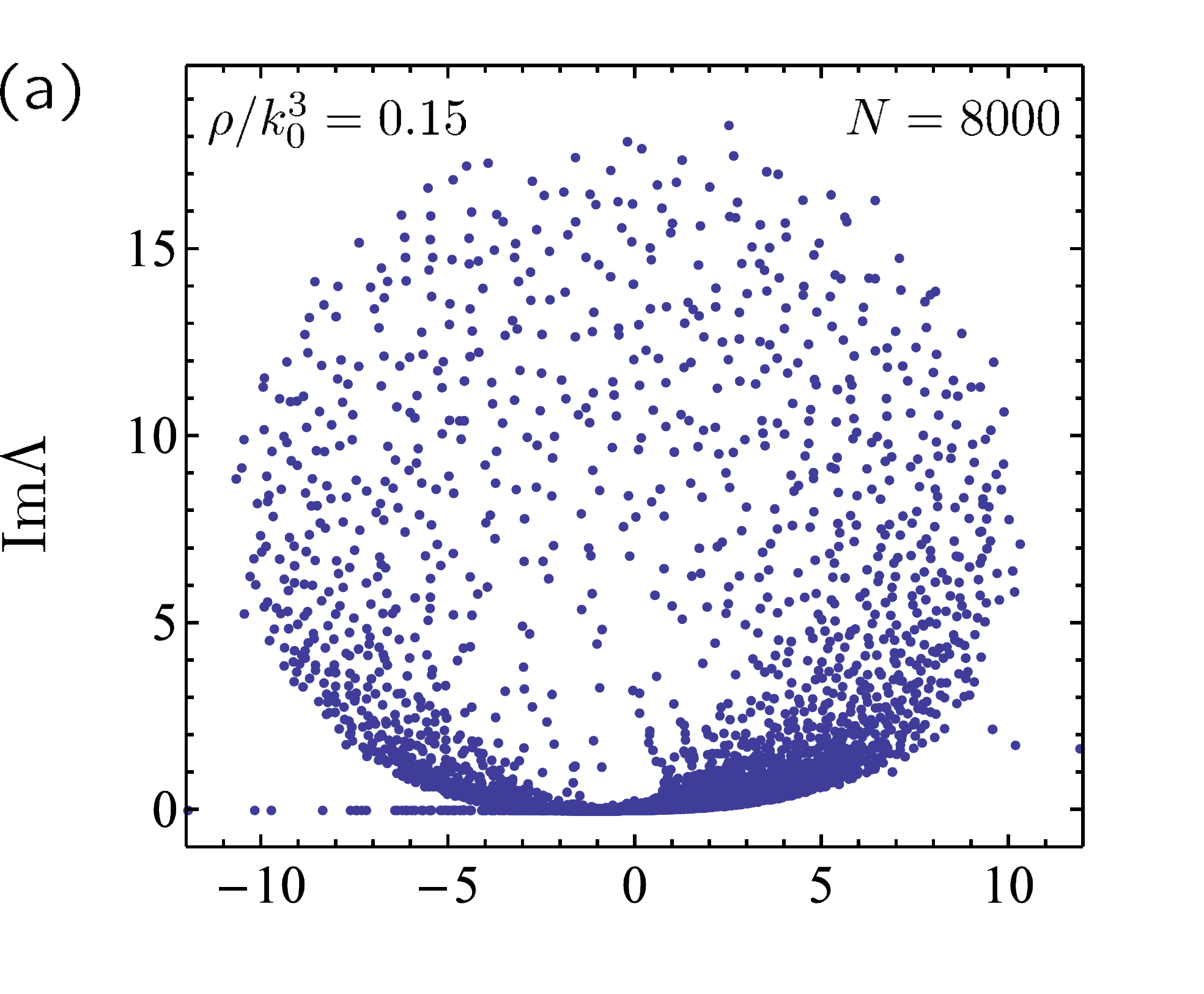}\\
\vspace{-6mm}
\includegraphics[width=0.97\columnwidth]{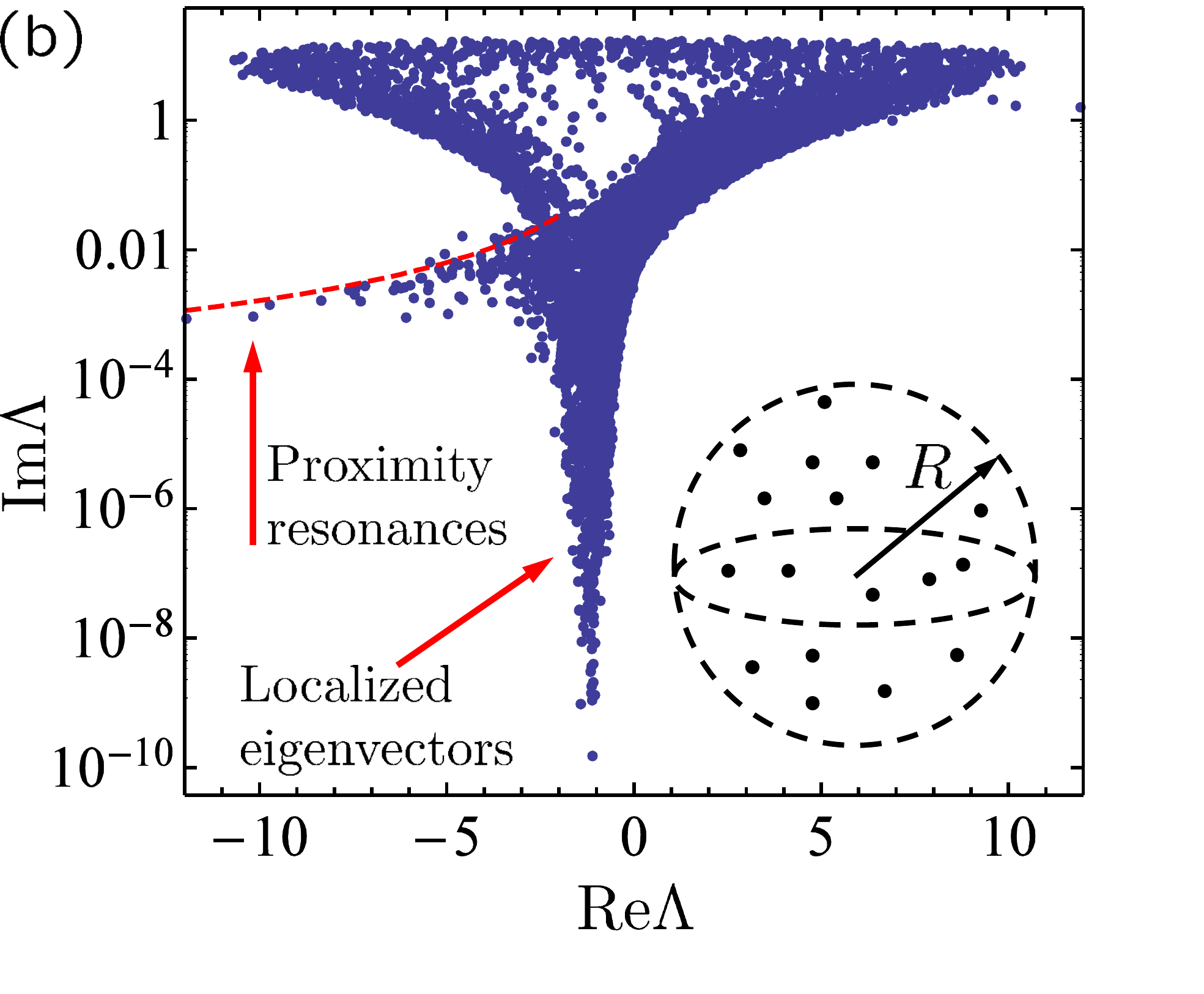}
\vspace{-3mm}
\caption{(a). Eigenvalues of a single random realization of the Green's matrix (\ref{green}) are shown by points on the complex plane for $N = 8000$ scatterers randomly distributed in a sphere with a density $\rho/k_0^3 = 0.15$. (b). Same as (a) but with the vertical axis in logarithmic scale. Arrows indicate eigenvalues corresponding to spatially localized eigenvectors and two-scatterer proximity resonances, respectively. The latter are concentrated along the dashed line given by Eq.\ (\ref{prox}). The inset illustrates the geometry, with points inside an imaginary sphere corresponding to scatterers at random positions $\{ \mathbf{r}_j \}$.}
\label{fig_illustration}
\end{figure}

\section{Green's matrix model}
\label{greenmodel}

A minimal realistic model in which Anderson localization of classical waves (such as sound, light, or elastic waves) can be efficiently studied considers a single excitation (a ``photon'' in the case of light) that propagates in an ensemble of $N$ point scattering centers (abstract ``point scatterers'', ``dipoles'' or ``atoms'' for light) randomly distributed in a volume $V$. Each scatterer is assumed strongly resonant (resonance frequency $\omega_0$, resonance width $\Gamma_0 \ll \omega_0$) and the delay due to the time needed for a wave to propagate through the medium without scattering is neglected: $L/c \ll 1/\Gamma_0$, where $L$ is the system size and  $c$ is the speed of the wave in the absence of scatterers. The quasi-modes of this system can be identified with eigenvectors of a non-Hermitian random Green's matrix ${\hat G}$ that for scalar waves in a 3D space is a $N \times N$ matrix with elements
\cite{skip14,rusek00,pinheiro04,skip11,goetschy11}
\begin{eqnarray}
G_{jk} = i \delta_{jk} + (1-\delta_{jk}) \frac{\exp(i k_0 |\vec{r}_j - \vec{r}_k|)}{k_0 |\vec{r}_j - \vec{r}_k|},
\label{green}
\end{eqnarray}
where $k_0 = \omega_0/c$ and $\left\{ \vec{r}_j \right\}$ are the positions of scatterers ($j = 1,\ldots,N$). Localization of quasi-modes in space can be studied not only by analyzing their spatial structure directly but also by examining the complex eigenvalues $\Lambda_n$ of ${\hat G}$. For the sake of illustration, we show $\Lambda_n$ of a random realization of ${\hat G}$ for points $\left\{ \vec{r}_j \right\}$ inside a sphere in Fig.\ \ref{fig_illustration}. The real part $\mathrm{Re} \Lambda_n$ of an eigenvalue $\Lambda_n$ yields the frequency $\omega_n = \omega_0 - (\Gamma_0/2) \mathrm{Re} \Lambda_n$ of the corresponding quasi-mode whereas its imaginary part $\mathrm{Im} \Lambda_n$ corresponds to the decay rate of the quasi-mode $\Gamma_n/2 = (\Gamma_0/2) \mathrm{Im} \Lambda_n$. A parameter analogous to the dimensionless (or Thouless) conductance \cite{abrahams79,edwards72,thouless77} can be defined as a ratio of $\Gamma_n/2$ to the average spacing $\langle \Delta \omega \rangle = \langle \omega_n - \omega_{n-1} \rangle$ between frequencies of quasi-modes in the vicinity of $\omega = \omega_n$:
\begin{eqnarray}
g_n = \frac{\Gamma_n/2}{\langle \Delta \omega \rangle} =
\frac{\mathrm{Im}\Lambda_n}{\langle \mathrm{Re}\Lambda_n - \mathrm{Re}\Lambda_{n-1} \rangle},
\label{gdef}
\end{eqnarray}
where the eigenvalues $\Lambda_n$ are ordered such that $\mathrm{Re}\Lambda_n > \mathrm{Re}\Lambda_{n-1}$. Defined in this way \cite{foot1}, Thouless conductance appears as nothing else than a normalized decay rate of an eigenmode. Obviously, $g_n$ is a random quantity and only its statistical properties are meaningful.

Statistical distributions of decay rates $\Gamma/2$ and normalized decay rates $g$ defined in a way analogous to ours have been previously studied for the tight-binding Hamiltonian of the open 3D Anderson model \cite{kottos02,weiss06}. It was shown that these distributions bear clear signatures of localization transition and that $p(g)$ takes a universal shape at the critical point, but a quantitative analysis allowing for estimating the critical exponents of the transition was not peformed. Here we apply a similar approach to the classical-wave system for which Eq.\ (\ref{green}) plays a role of an effective Hamiltonian and perform a finite-size scaling analysis that yields an estimate of the critical exponent $\nu$ of the localization transition in our model. Although the same model has been studied by several authors \cite{skip14,rusek00,pinheiro04,bellando14}, no proper finite-size scaling has been realized and no estimation of $\nu$ has been proposed up to now.

We have previously shown \cite{skip14} that the matrix (\ref{green}) starts to have spatially localized eigenvectors when the number density of scatterers $\rho=N/V$ exceeds $\rho \approx 0.1 k_0^3$ and that all eigenvectors become again extended beyond $\rho \approx k_0^3$. Localized eigenvectors correspond to eigenvalues $\Lambda$ with a very small imaginary part and the real part in a narrow spectral range around $\mathrm{Re} \Lambda \approx -1$, see Fig.\ \ref{fig_illustration}(b). Instead of studying the localization transition for a given $\mathrm{Re} \Lambda$ as a function of increasing density as in Ref.\ \onlinecite{skip14}, one can also keep the density constant (though sufficiently high) and study the transition as a function of $\mathrm{Re} \Lambda$ or, equivalently, frequency $\omega = \omega_0 - (\Gamma_0/2) \mathrm{Re} \Lambda$, which may be closer to realistic experimental scenarios. This is the approach that we follow in the present work.
We will refer to $\mathrm{Re} \Lambda$ as ``frequency" for short.

In addition to the localized states appearing at high densities of scatterers, so-called subradiant or ``dark'' states corresponding to proximity resonances of closely located pairs of scatterers \cite{heller96,rusek00} exist at any density \cite{skip11,goetschy11}. A line along which eigenvalues corresponding to proximity resonances are concentrated can be found by diagonalizing a $2 \times 2$ Green's matrix:
\begin{eqnarray}
\Lambda = -\frac{\cos(k_0 \Delta r)}{k_0 \Delta r} - i \left[ \frac{\sin(k_0 \Delta r)}{k_0 \Delta r} - 1 \right],
\label{prox}
\end{eqnarray}
where $\Delta r \ll 1/k_0$ is the distance between the two scatterers on which the resonant state is localized. These states have large negative frequency shifts $\mathrm{Re} \Lambda \ll -1$ and their eigenvalues belong to the ``tail'' on the left from the main eigenvalue cloud in Fig.\ \ref{fig_illustration}. Large frequency shifts allow for discriminating proximity resonances from the states localized due to disorder in the system. Therefore, these two types of localized states can be analyzed separately by choosing a particular range of $\mathrm{Re} \Lambda$.

\begin{figure}
\includegraphics[width=0.92\columnwidth]{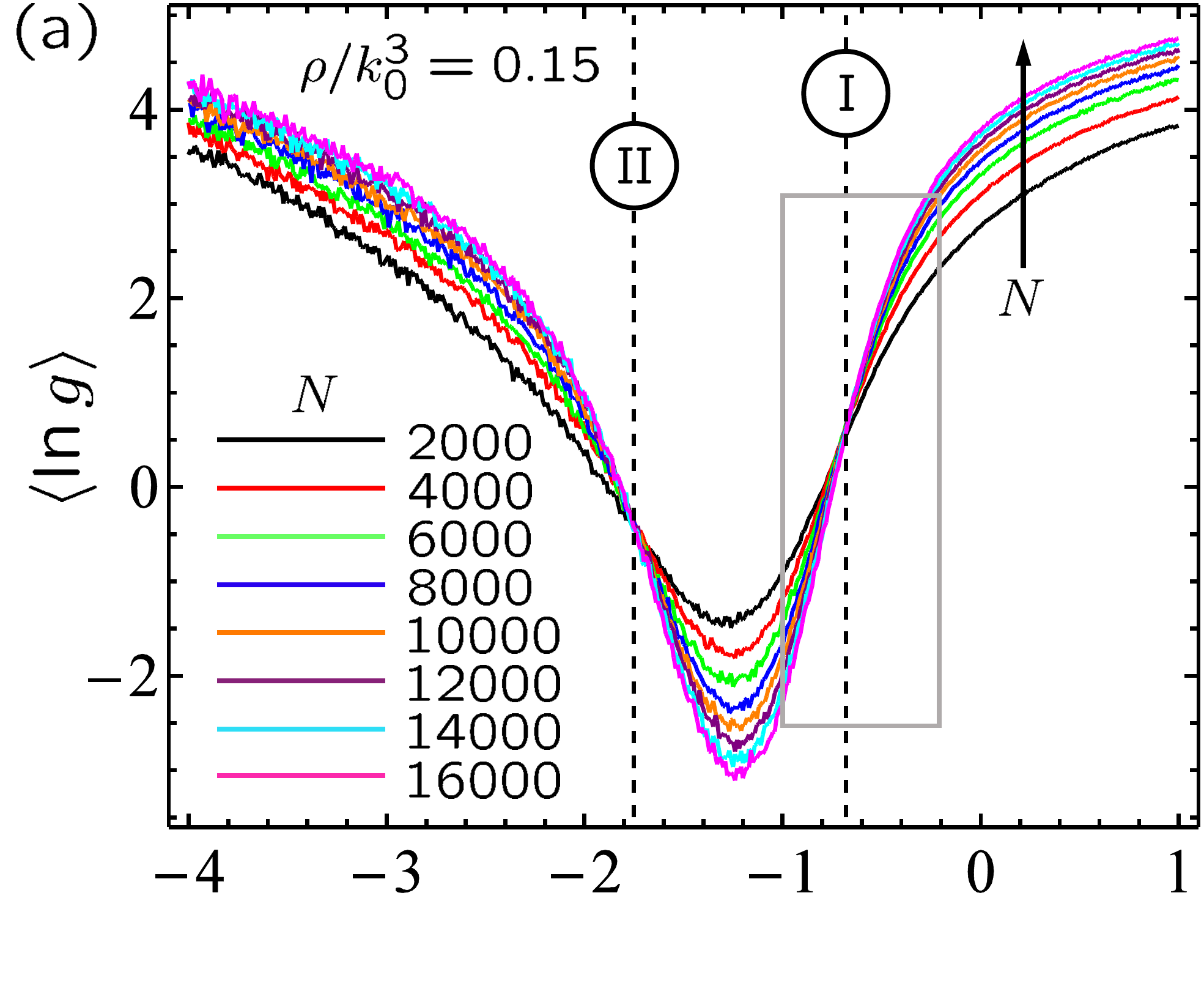}\\
\vspace{-6mm}
\includegraphics[width=0.92\columnwidth]{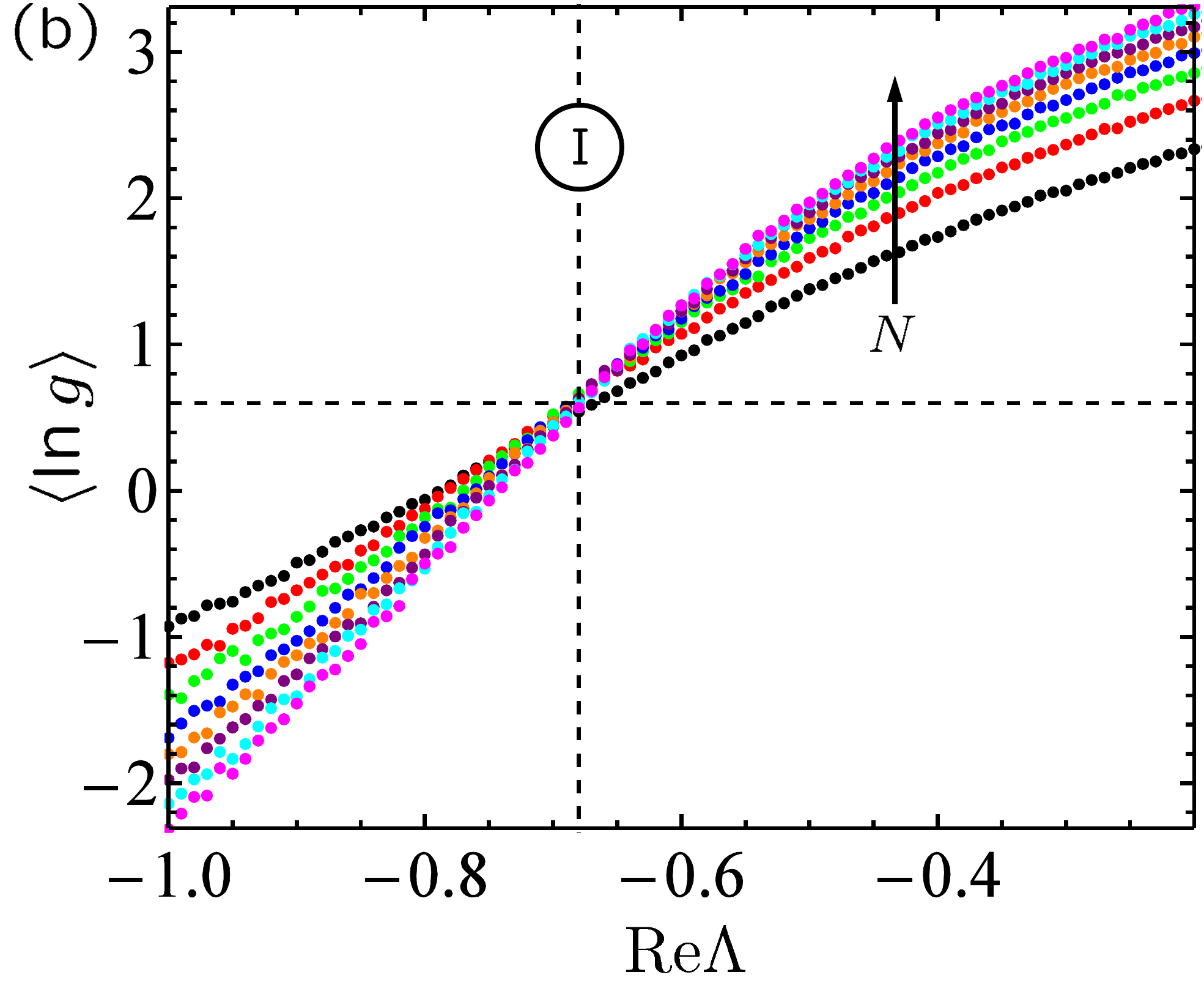}
\caption{(a). Average logarithm of the Thouless conductance as a function of frequency $\mathrm{Re} \Lambda$ at a given number density $\rho/k_0^3 = 0.15$ and for several total numbers $N$ of scatterers from $N = 2000$ to $N = 16000$ (the order of curves corresponding to increasing $N$ is indicated by an arrow). Vertical dashed lines show the estimated locations of the two mobility edges (denoted as I and II) where curves corresponding to different $N$ all cross. (b). Zoom on the region around the mobility edge I, shown by a grey rectangle in panel (a). These results are obtained by averaging over 6385, 3140, 1798, 1250, 1120, 880, 744 and 665 numerically generated and diagonalized \cite{lapack} realizations of random Green's matrices for $N$ from 2000 to 16000, respectively.}
\label{fig_lng_average}
\end{figure}

\section{Statistics of normalized decay rates}
\label{stat}

\begin{figure}
\includegraphics[width=0.95\columnwidth]{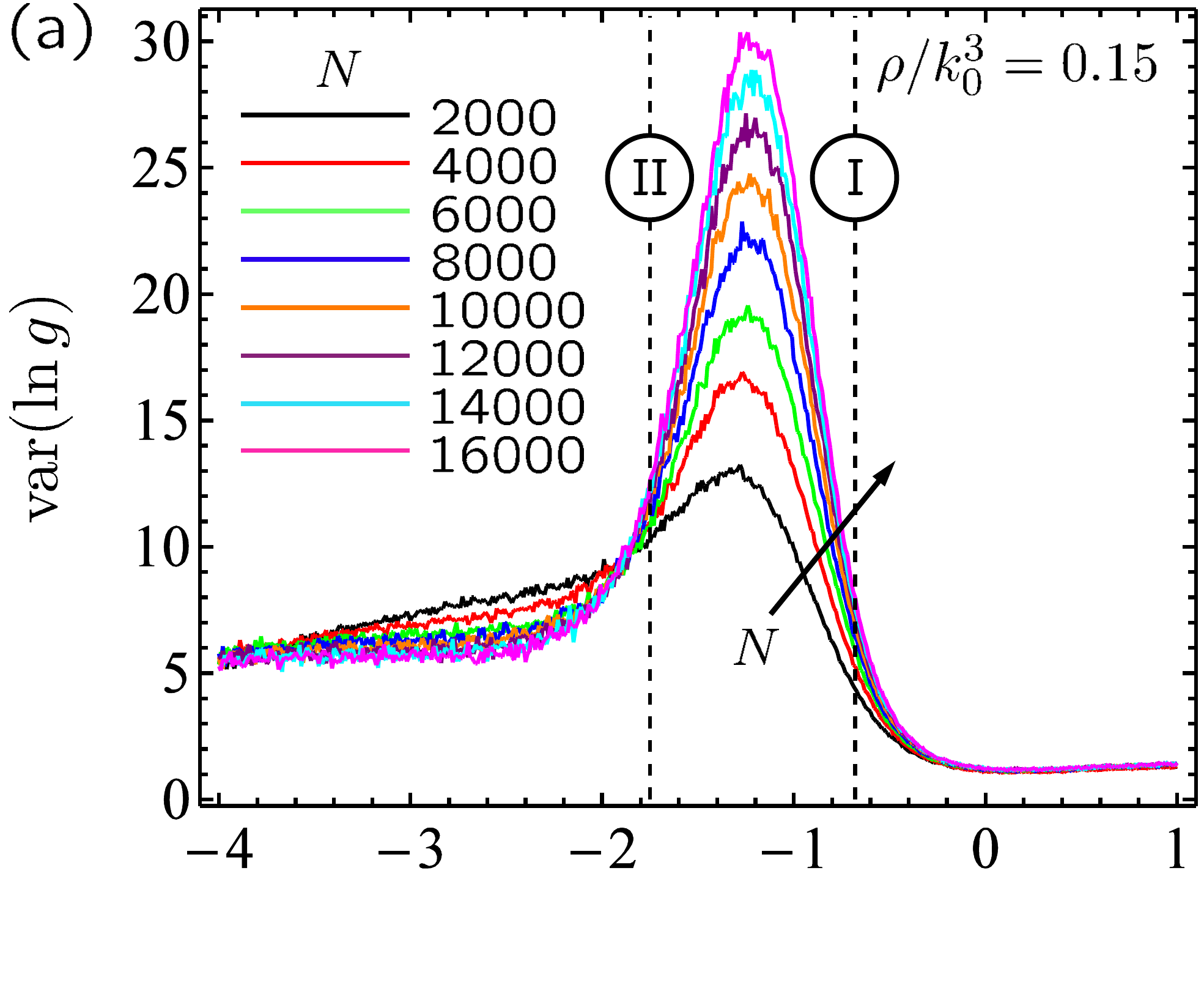}\\
\vspace{-7mm}
\includegraphics[width=0.95\columnwidth]{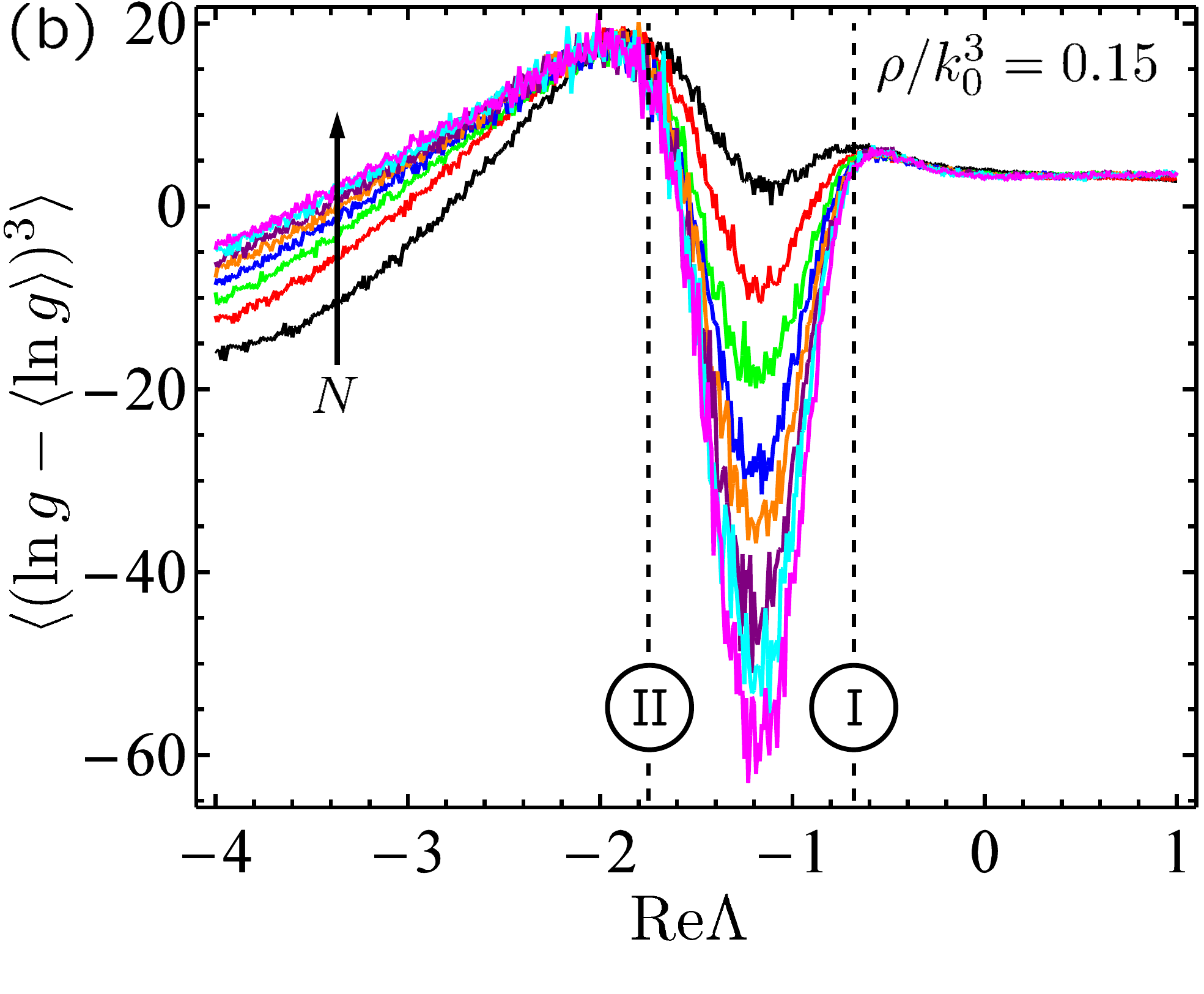}\\
\vspace{-3mm}
\caption{The variance (a) and the third central moment (b) of $\ln g$. The order of curves corresponding to increasing $N$ is indicated by arrows. Vertical dashed lines are the same as in Fig.\ \ref{fig_lng_average}(a).}
\label{fig_lng_moments}
\end{figure}

Following Ref.\ \onlinecite{skip14}, we consider $N$ resonant point scattering centers randomly distributed in a spherical volume of radius $R$, see the inset of Fig.\ \ref{fig_illustration}(b). The $N \times N$ Green's matrix (\ref{green}) describing the propagation of scalar waves between the scatterers is diagonalized numerically for many different scatterer configurations $\{\vec{r}_j\}$ (see Fig.\ \ref{fig_illustration} for an example of eigenvalues obtained for a single configuration) and the statistics of the Thouless conductance $g$ defined by Eq.\ (\ref{gdef}) is studied. We perform calculations at a fixed scatterer density $\rho/k_0^3 = 0.15$ which is high enough to ensure appearance of localized states \cite{skip14}. The number of scatterers $N$ is varied in a range $N = 2000$--16000 with the number of independent configurations $\{ \mathbf{r}_j \}$ adjusted to ensure a total of at least $10^7$ eigenvalues $\Lambda$ for each $N$. The results are analyzed as a function of $\mathrm{Re} \Lambda$. For a given value of $\mathrm{Re} \Lambda$, averaging denoted by $\langle \cdots \rangle$ is performed over the realizations of disorder $\{\vec{r}_j\}$ and over a narrow frequency interval $\mathrm{Re} \Lambda \pm \frac12 \delta_{\mathrm{Re} \Lambda}$ with $\delta_{\mathrm{Re} \Lambda} = 0.01$.

\begin{figure}
\includegraphics[width=0.95\columnwidth]{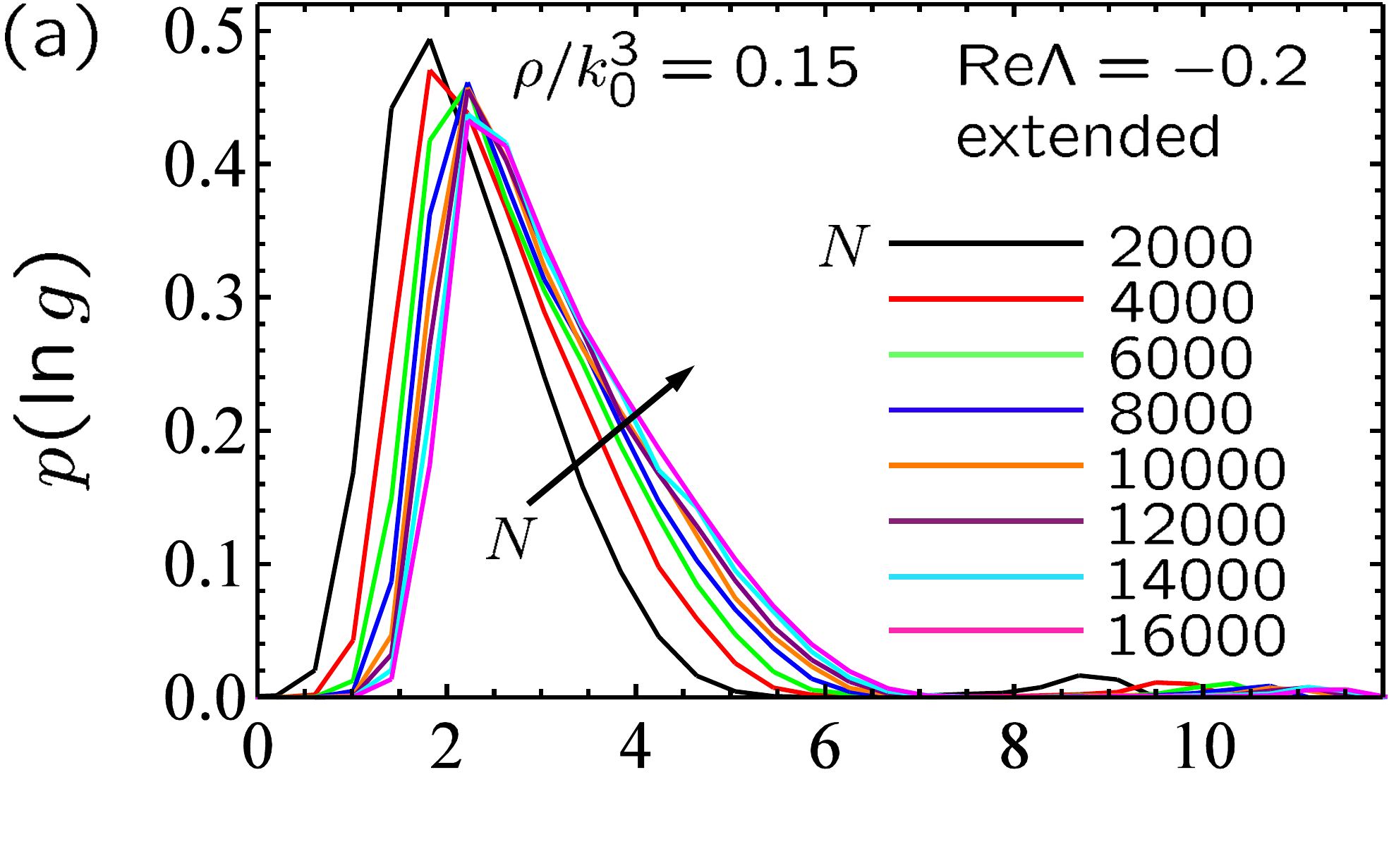}\\
\vspace*{-4mm}
\includegraphics[width=0.95\columnwidth]{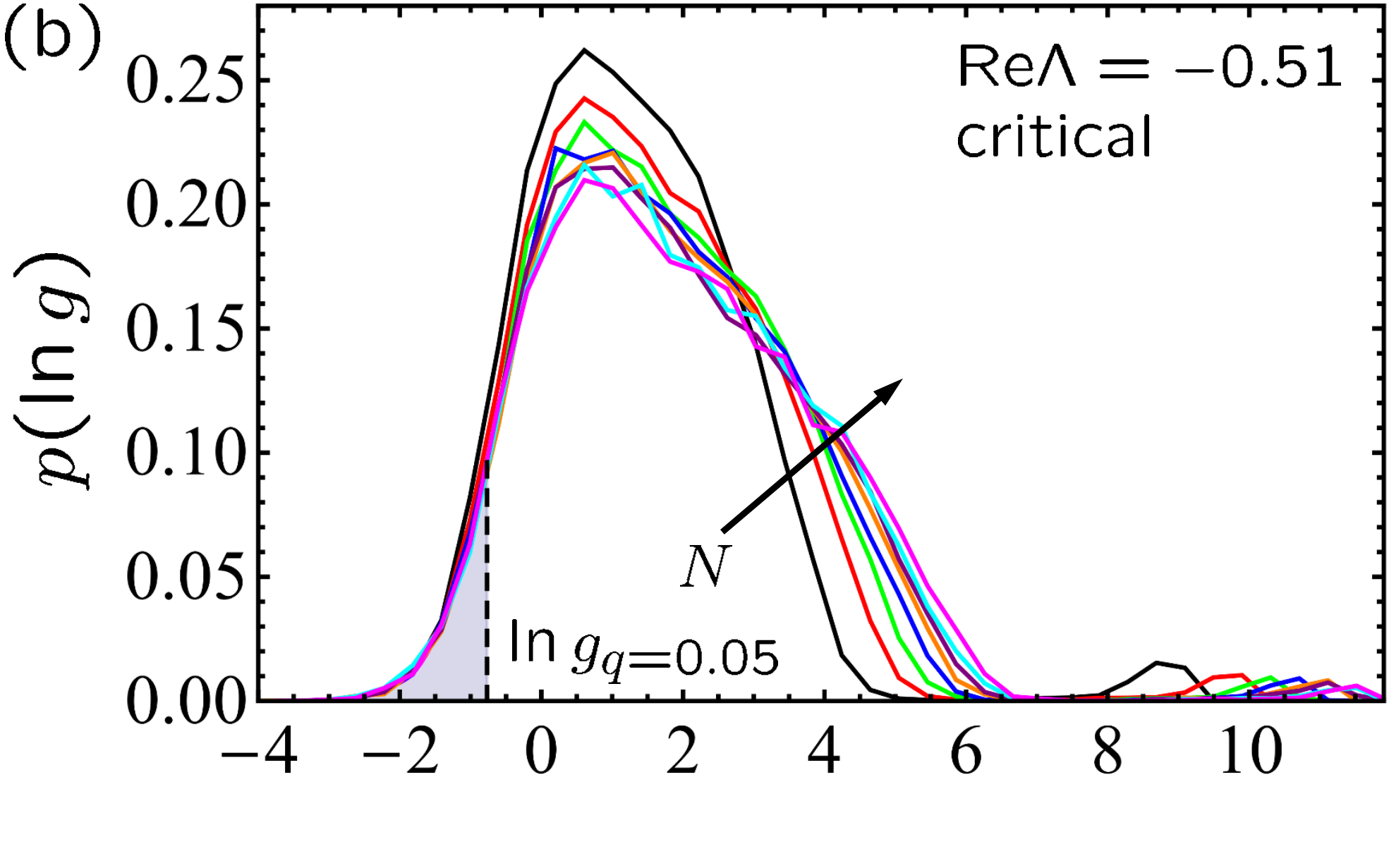}\\
\vspace*{-4mm}
\includegraphics[width=0.95\columnwidth]{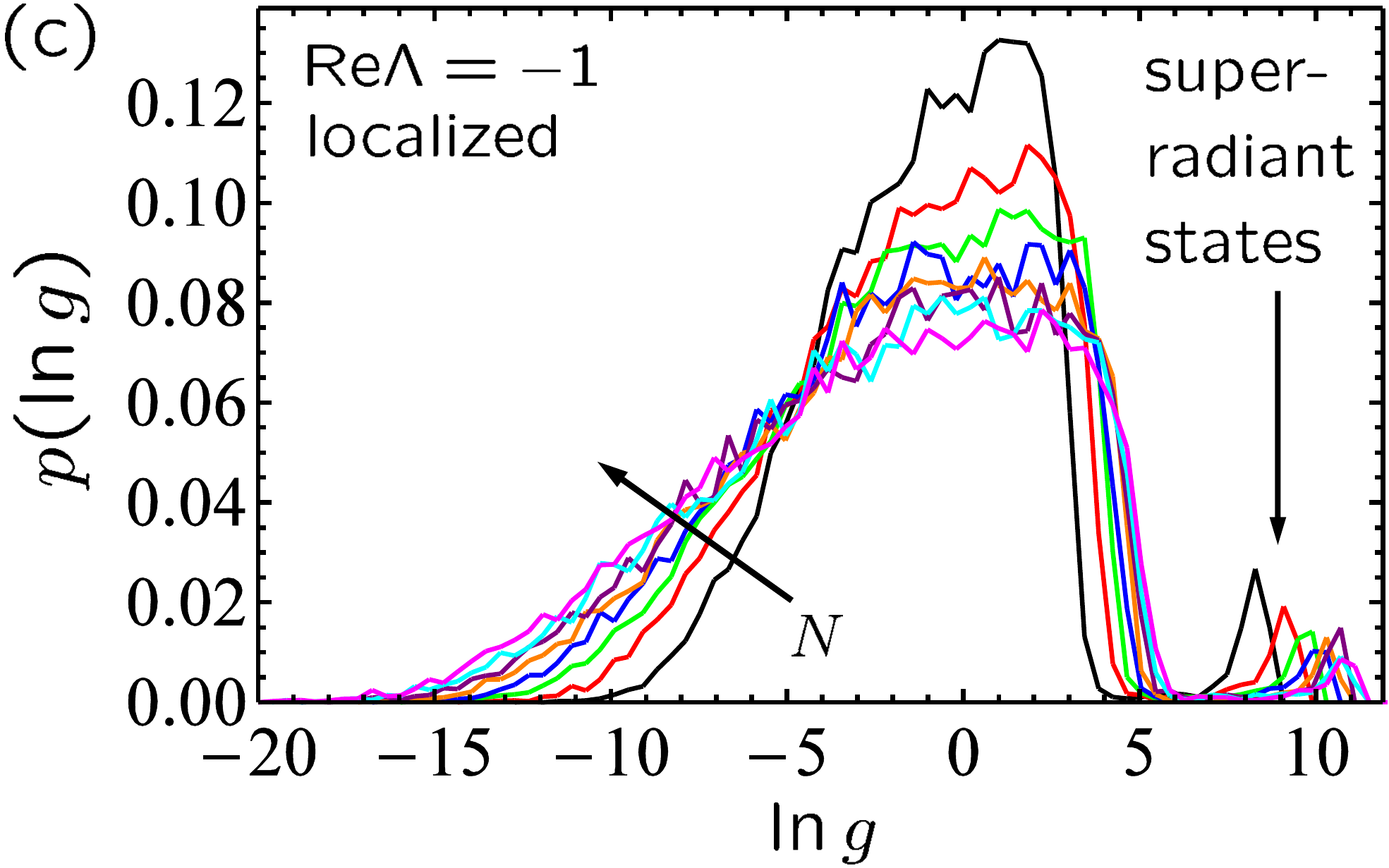}
\caption{Evolution of the probability density of $\ln g$ from the extended regime (a) through the critical point (b) to a regime in which localized states appear (c), in the vicinity of the localization transition I. The order of curves corresponding to increasing $N$ is indicated by arrows. Dashed line in (b) shows the percentile $g_q$ for $q=0.05$. Distributions at the critical point can be considered independent of $N$ on the left from the dashed line.}
\label{fig_lng_distr}
\end{figure}

\begin{figure}
\includegraphics[width=0.95\columnwidth]{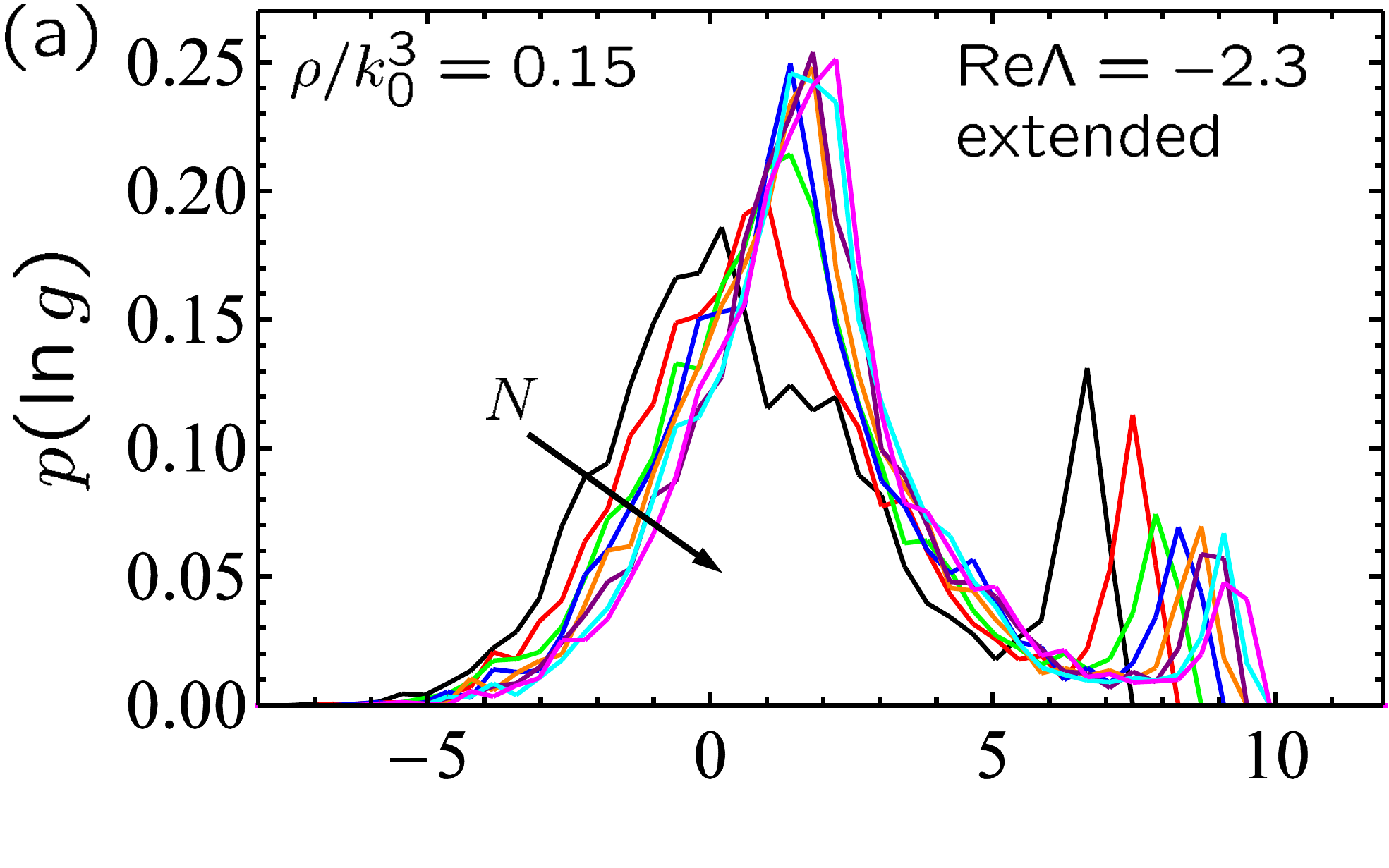}\\
\vspace*{-4mm}
\includegraphics[width=0.95\columnwidth]{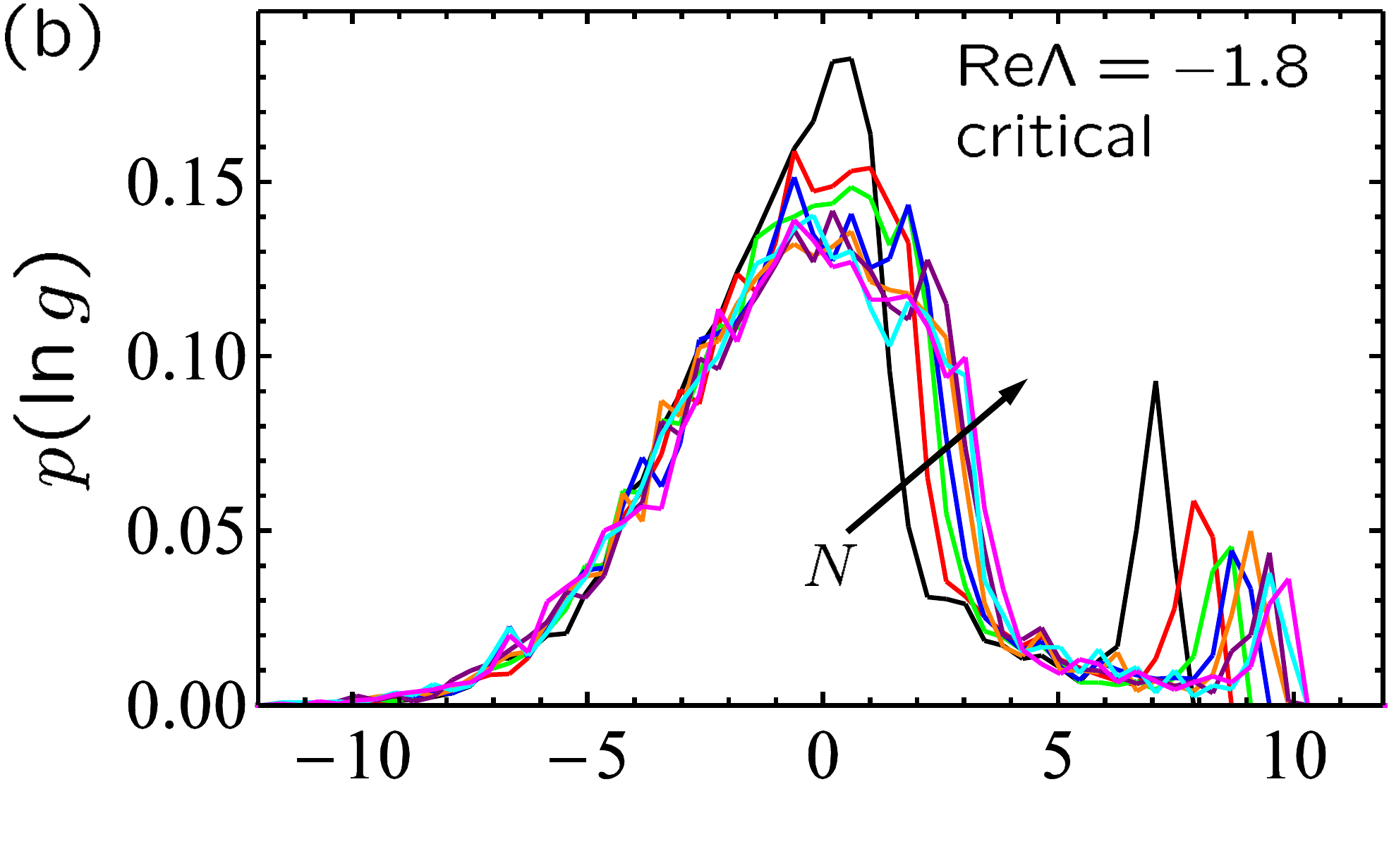}\\
\vspace*{-4mm}
\includegraphics[width=0.95\columnwidth]{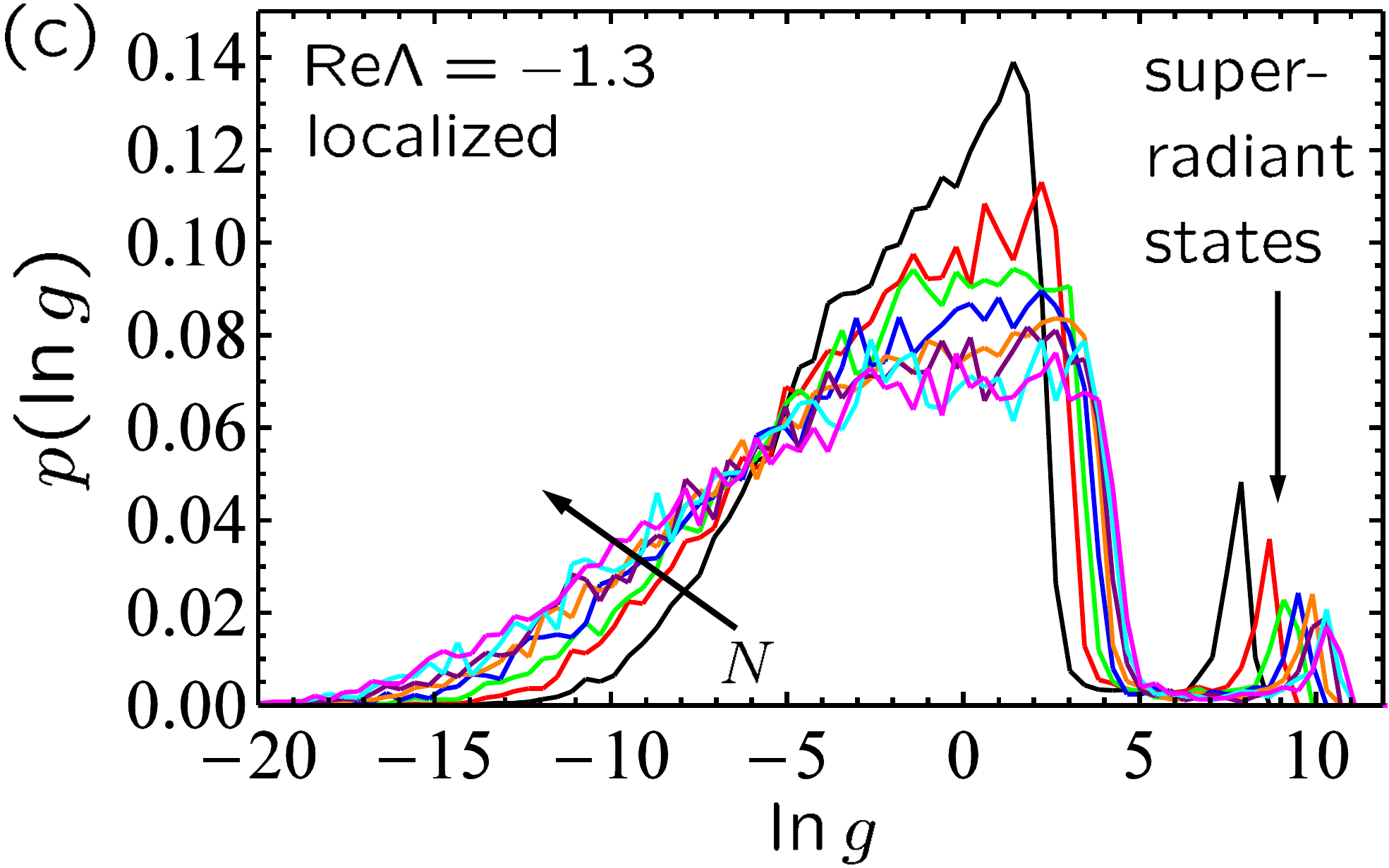}
\caption{Same as Fig.\ \ref{fig_lng_distr} but for the localization transition II.}
\label{fig_lng_distr_2nd}
\end{figure}

Figure \ref{fig_lng_average} shows the average logarithm of $g$ as a function of frequency $\mathrm{Re} \Lambda$. The curves corresponding to different system sizes $N$ all cross in two points; the abscissas of these points can be taken as rough estimates of positions of the two mobility edges that we denote by I and II, respectively. The frequency range between the mobility edges roughly corresponds to a band of localized quasi-modes (or, equivalently, a mobility gap) discovered in our previous work \cite{skip14}. The existence of a mobility gap between two mobility edges instead of a single mobility edge separating extended states at high energies (frequencies) from localized states at low energies (frequencies) is typical for resonant scattering \cite{cobus15}. It may be tempting to use the numerical data of Fig.\ \ref{fig_lng_average} to perform the standard finite-size scaling analysis (for example, along the lines of Ref.\ \onlinecite{slevin01}) in order to estimate such parameters as the precise locations of mobility edges and the value of the critical exponent. One, however, should be careful because it is known that the distribution of $g$ and even of $\ln g$ can be very broad at the transition point and, as a consequence, statistical moments of $g$ and $\ln g$ may be dominated by nonuniversal tails of the distribution that are not expected to obey any scaling laws \cite{shapiro86,cohen88}. This is illustrated by Figs.\ \ref{fig_lng_moments}, \ref{fig_lng_distr} and \ref{fig_lng_distr_2nd} where we show the variance and the third central moment of $\ln g$ and the full probability density $p(\ln g)$. Figure \ref{fig_lng_moments} shows, in particular, that $\mathrm{var}(\ln g) = \langle (\ln g)^2 \rangle - \langle \ln g \rangle^2$ corresponding to different $N$ has no crossing point around the estimated mobility edge I. At the same time, a crossing point existing near the mobility edge II does not coincide with the latter. The third central moment may exhibit crossing points near (although not exactly at) expected mobility edges but they are masked by the statistical noise in the data. As a consequence, simple relations between moments of $\ln g$ found for the Anderson model with diagonal disorder \cite{somoza06, somoza07} does not hold in our case. All this indicates that extracting the critical parameters of the localization transition in our system from the analysis of moments of $\ln g$ is not possible because moments of different orders would yield different results or no result at all as the variance, for example. The analysis of probability densities $p(\ln g)$ illustrates the reason behind this. Figures \ref{fig_lng_distr} and \ref{fig_lng_distr_2nd}  show $p(\ln g)$ in three different regimes: the extended regime (a), the vicinity of the critical point (b), and the localized regime (c). We note the broadness of distributions at the critical point and in the localized regime and superradiant peaks of $p(\ln g)$ at large $\ln g$. The superradiant states originate from collective effects and constitute a distinctive feature of dense ensembles of resonant point scatterers \cite{dicke54,gross82,svid10}. Although the phenomenon of superradiance is different from Anderson localization, it can disturb the analysis of the latter by, for example, yielding sizable contributions to the statistical moments of $\ln g$.

When studying the evolution of $p(\ln g)$ from Fig.\ \ref{fig_lng_distr}(a) to Fig.\ \ref{fig_lng_distr}(c) with decreasing $\mathrm{Re} \Lambda$ and from Fig.\ \ref{fig_lng_distr_2nd}(a) to Fig.\ \ref{fig_lng_distr_2nd}(c) with increasing $\mathrm{Re} \Lambda$, we did not find a value of $\mathrm{Re} \Lambda$ for which the probability densities corresponding to different $N$ would all coincide with each other as could be expected at a critical point. This is the reason behind the fact that there is no value of $\mathrm{Re} \Lambda$ at which moments of $\ln g$ would all be independent of $N$, as can be seen from Figs.\ \ref{fig_lng_average} and \ref{fig_lng_moments}. This also means that, strictly speaking, $p(g)$ does not obey a single-parameter scaling law. However, it follows from our analysis that the small-$g$ part of $p(\ln g)$ becomes roughly independent of $N$ around $\mathrm{Re} \Lambda \simeq -0.51$ [Fig.\ \ref{fig_lng_distr}(b)] and $\mathrm{Re} \Lambda \simeq -1.8$ [Fig.\ \ref{fig_lng_distr_2nd}(b)]. Because small $g$ correspond to long-lived quasi-modes that become spatially localized [see Fig.\ \ref{fig_illustration}(b)], we associate the universal, $N$-independent shape of $p(\ln g)$ for small $g$ with the critical points of localization transitions. The above values of $\mathrm{Re} \Lambda$ can be taken as new, improved estimates of positions of mobility edges I and II, respectively.  In contrast, the non-universal, $N$-dependent behavior of $p(\ln g)$ for large $g$ is due to extended states (including the superradiant states indicated by arrows in Figs.\ \ref{fig_lng_distr}(c) and \ref{fig_lng_distr_2nd}(c), but not only) that still exist in the system for any finite $N$. Figures \ref{fig_lng_distr} and \ref{fig_lng_distr_2nd} suggest that the statistical weight of these states seems to decrease with $N$ (the height of the peak corresponding to superradiant states decreases with $N$, for example), so that one can expect them to become statistically irrelevant in the limit of $N \to \infty$. However, the results in hand do not allow us to claim this with certainty.

\begin{figure}
\includegraphics[width=0.95\columnwidth]{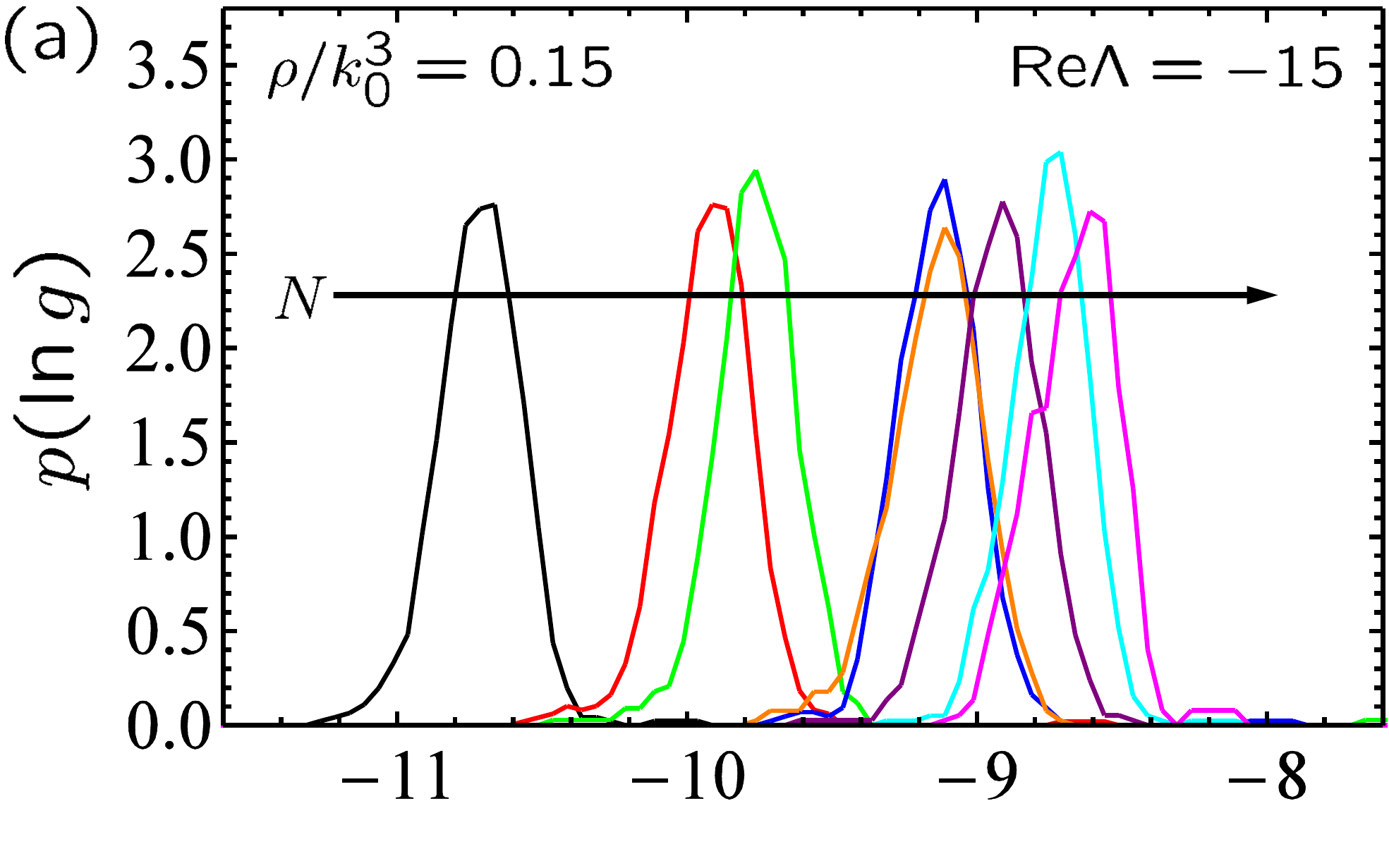}\\
\vspace*{-3.5mm}
\includegraphics[width=0.95\columnwidth]{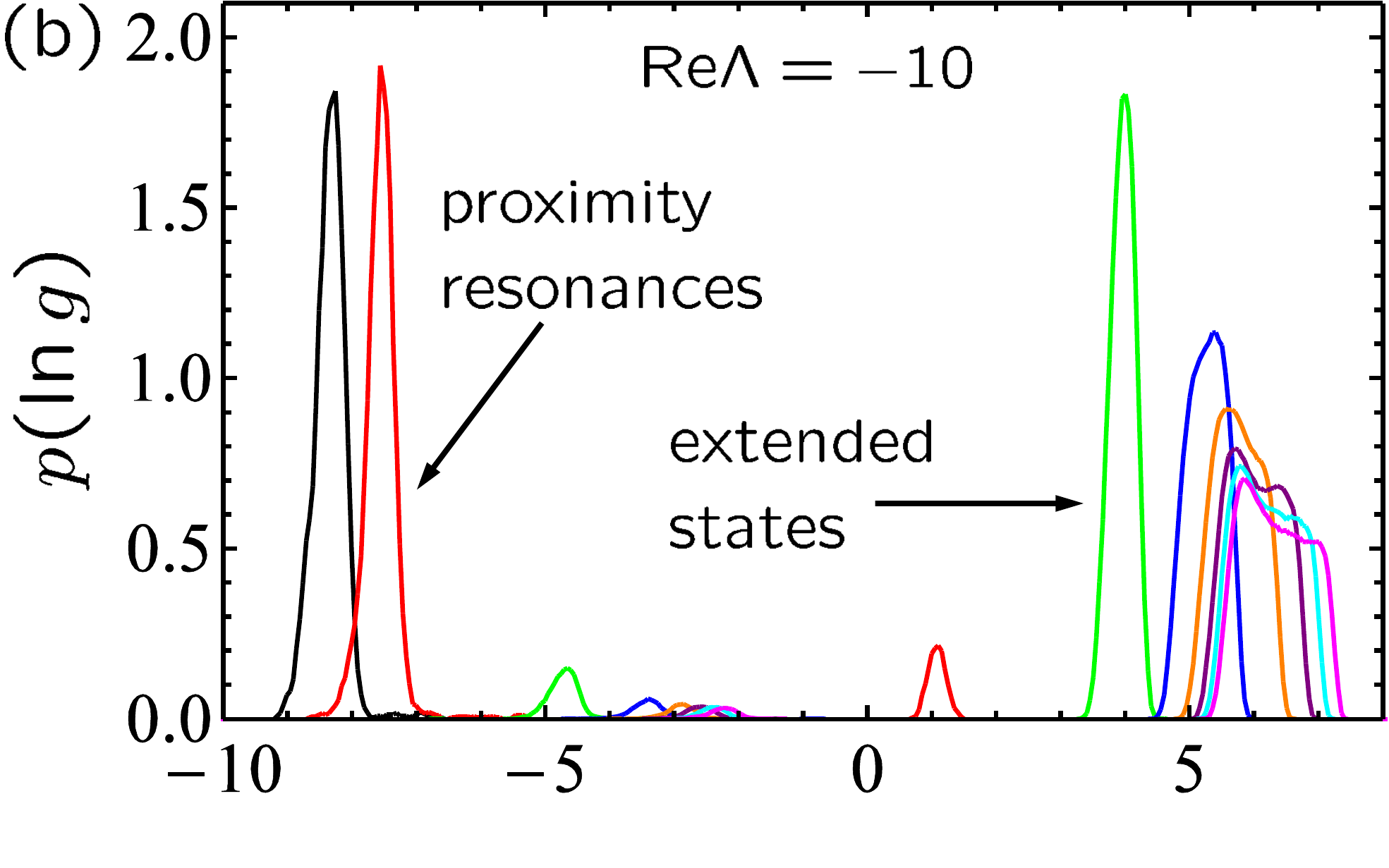}\\
\vspace*{-4mm}
\includegraphics[width=0.95\columnwidth]{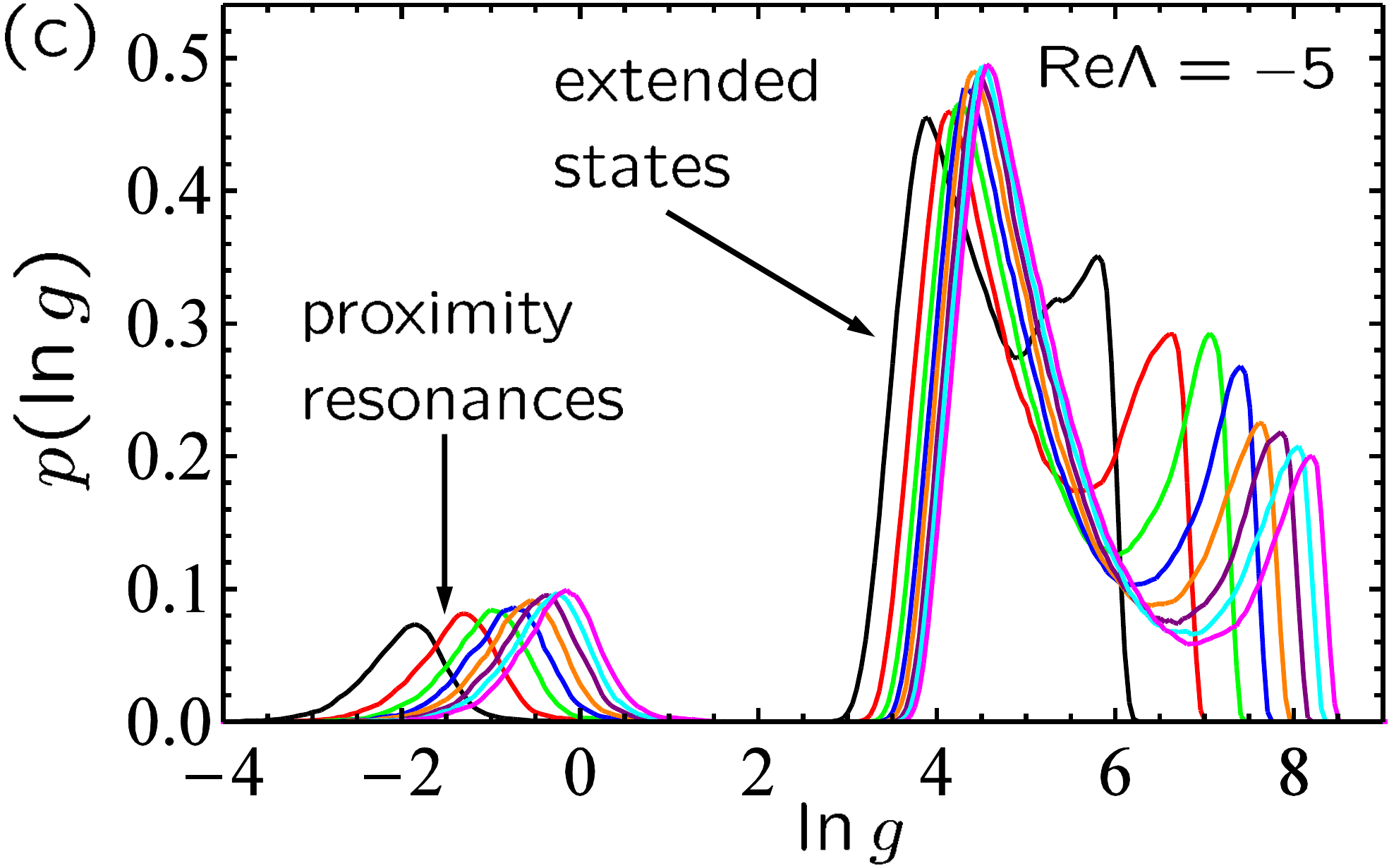}
\caption{Probability density of $\ln g$ at large negative $\mathrm{Re} \Lambda$ where subradiant states localized on pairs of scatterers (proximity resonances) come into play. To improve statistics, averaging is performed over a range $\delta_{\mathrm{Re} \Lambda} = 1$ around the values of $\mathrm{Re} \Lambda$ given on the plots instead of $\delta_{\mathrm{Re} \Lambda} = 0.01$ in Figs.\ \ref{fig_lng_distr} and \ref{fig_lng_distr_2nd}. This is made possible by the slow dependence of the properties of proximity resonances on $\mathrm{Re} \Lambda$.}
\label{fig_lng_distr_subrad}
\end{figure}

For completeness, we close this section by considering $p(\ln g)$ in a frequency range where subradiant states localized on pairs of closely located scatterers (i.e., proximity resonances) play an important role. The latter are the only states present at very large negative $\mathrm{Re} \Lambda$. They give narrow, peaked distributions of $\ln g$ shown in Fig.\ \ref{fig_lng_distr_subrad}(a). As a rule, $p(\ln g)$ shifts to the right when increasing $N$ which might appear counterintuitive because the states under consideration are localized and, by analogy with Anderson localization, one naively expects typical values of $\ln g$ to decrease with system size and hence with $N$. Here, however, the mechanism of localization is different from the Anderson one and the above analogy does not apply. At a given $\mathrm{Re} \Lambda$, the value of the numerator $\mathrm{Im} \Lambda$ in the definition (\ref{gdef}) of $g$ follows from Eq.\ (\ref{prox}) which is independent of $N$. The denominator, however, decreases with $N$ simply because spacings between eigenvalues become smaller when the number of eigenvalues $N$ grows. As a result, typical values of $g$ increase with $N$ as we see in Fig.\ \ref{fig_lng_distr_subrad}(a).

When $\mathrm{Re} \Lambda$ is increased, extended states from the ``bulk'' of eigenvalue cloud (see Fig.\ \ref{fig_illustration}) start to contribute to $p(\ln g)$, first for large $N$ only [see Fig.\ \ref{fig_lng_distr_subrad}(b)] and then for all $N$ [Fig.\ \ref{fig_lng_distr_subrad}(c)]. Note, however, that larger $N$ always correspond to larger typical (i.e., average) values of $\ln g$ as follows from Fig.\ \ref{fig_lng_distr_subrad}(a--c) where $p(\ln g)$ has a clear tendency to shift to the right with increasing $N$. We thus conclude that the two-scatterer proximity resonances exhibit neither the scaling with system size expected for Anderson localization nor any signatures of critical behavior. We will not consider the part of the eigenvalue spectrum  $\mathrm{Re} \Lambda \ll -1$ corresponding to these resonances in the remainder of the paper.

\section{Scaling theory and single-parameter scaling of percentiles}
\label{sps}

\begin{figure*}
\includegraphics[width=0.95\textwidth]{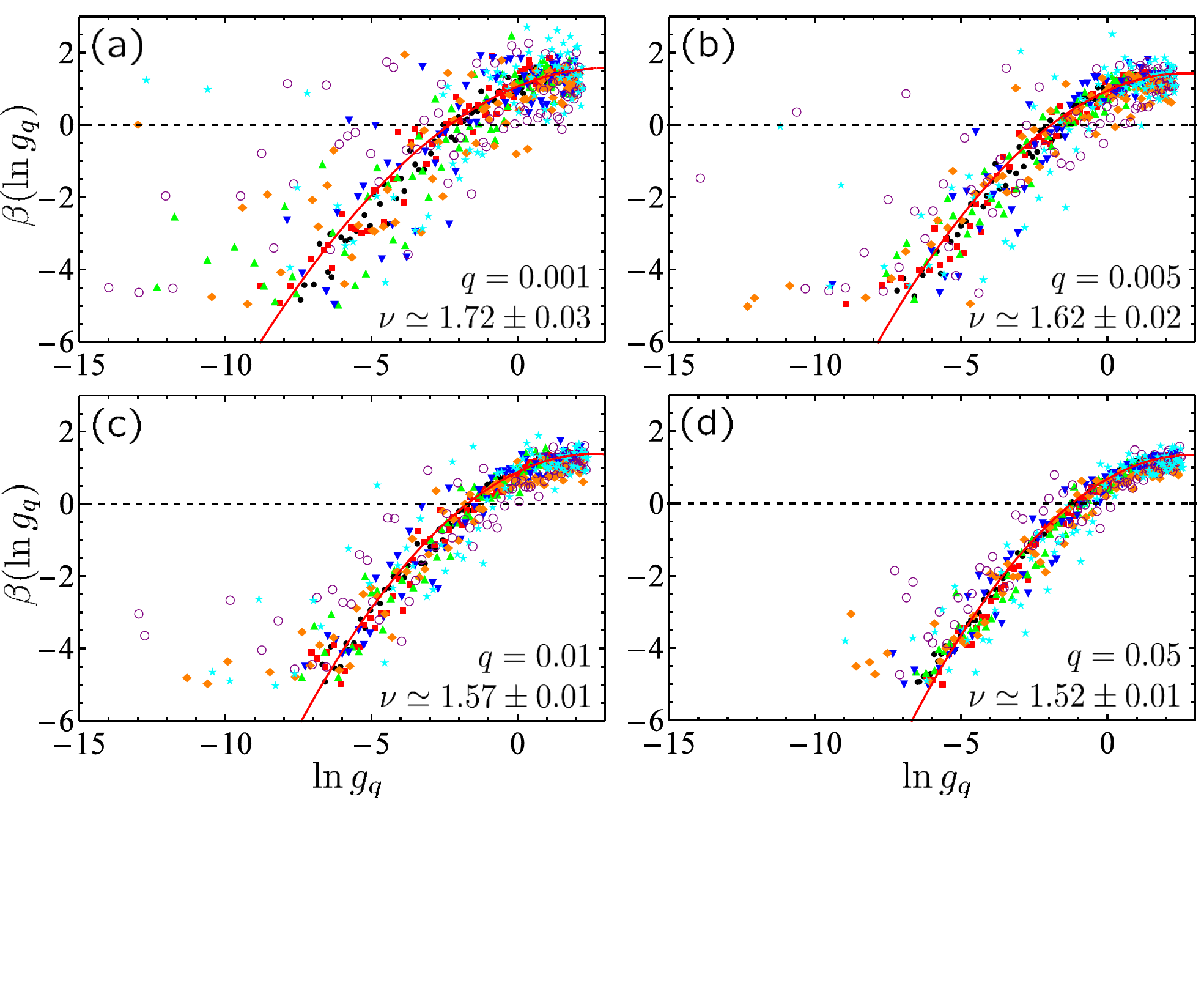}
\vspace*{-3cm}
\caption{Examples of second-order polynomial fits (solid lines) to the numerical data (symbols) for the $\beta$-function $\beta(\ln g_q)$. Fits for four different values of $q$ are shown. Only data points corresponding to $\beta > -5$ were used for fitting and are shown in the figures. Different symbols correspond to estimations of the $\beta$-function for
$N = 2$ and $4 \times 10^3$ (full circles),
$N = 4$ and $6 \times 10^3$ (squares),
$N = 6$ and $8 \times 10^3$ (triangles),
$N = 8$ and $10 \times 10^3$ (upside-down triangles),
$N = 10$ and $12 \times 10^3$ (diamonds),
$N = 12$ and $14 \times 10^3$ (open circles),
$N = 14$ and $16 \times 10^3$ (stars).
Values of the critical exponent $\nu$ following from the fits are shown on the figures.
}
\label{fig_beta}
\end{figure*}

To avoid the impact of the large-$g$ nonuniversal part of $p(g)$ visible in Figs.\ \ref{fig_lng_distr} and \ref{fig_lng_distr_2nd} on our analysis, we consider percentiles $g_q$ defined by the following equality:
\begin{eqnarray}
q = \int\limits_{0}^{g_q} p(g) dg.
\label{perc}
\end{eqnarray}
The definition of a percentile is illustrated in Fig.\ \ref{fig_lng_distr}(b) where the shaded area is equal to $q = 0.05$ and the vertical dashed line shows $\ln g_q$. According to Eq.\ (\ref{perc}), the region of $\ln g_q$ that counts for the calculation of $g_{q = 0.05}$ is on the left from the dashed line in Fig.\ \ref{fig_lng_distr}(b) \cite{foot3}.
Small-$q$ percentiles depend only on the small-$g$ part of the distribution $p(g)$ and are therefore suitable for the analysis of the localization transition. Scaling analysis of $g_q$ has been previously used to demonstrate the single-parameter scaling of conductance distribution in the Anderson model \cite{slevin2003}. Here we analyze $g_q$ for $q = 0.001$--$0.05$ and restrict ourselves to the vicinity of the localization transition I. The transition II can, in principle, be analyzed in the same way but it takes place in a spectral region where the eigenvalue density of the Green's matrix ${\hat G}$ is lower, requiring a larger number of independent scatterer configurations $\{ \mathbf{r}_j \}$ to reach an acceptable statistical accuracy. Consequences of this can be seen, for example, in Figs.\ \ref{fig_lng_average}(a) and \ref{fig_lng_moments} where the estimated moments of $\ln g$ exhibit significantly stronger statistical fluctuations in the vicinity of transition II than in the vicinity of transition I. In addition, it follows from the comparison of Figs.\ \ref{fig_lng_distr}(b) and \ref{fig_lng_distr_2nd}(b) that the relative statistical weight of superradiant states is larger at the transition II, making this transition less suitable for an accurate scaling analysis.

The scaling theory of localization \cite{abrahams79} teaches us that a ``typical'' or ``scaling'' conductance $\tilde{g}$ is expected to obey a scaling law
\begin{eqnarray}
\frac{\partial \ln \tilde{g}(L)}{\partial \ln L} = \beta(\ln \tilde{g}).
\label{beta}
\end{eqnarray}
Thus, $\tilde{g}$ (or, equivalently, $\ln \tilde{g}$) is the only relevant scaling variable.
One expects $\beta > 0$ (conductance grows with system size $L$) when eigenstates are extended and transport is diffusive, $\beta < 0$ (conductance decreases with $L$) when eigenstates are localized, and $\beta = 0$ (conductance independent of $L$) at the mobility edge. It has been noticed that the average conductance $\langle g \rangle$ generally cannot be taken as $\tilde{g}$ because it can be dominated by irrelevant tails of $p(g)$. Some ``representative'' conductance, such as, e.g., the median conductance $g_{q=0.5}$, should be considered instead \cite{shapiro86,cohen88}. Here we will assume that $\tilde{g} = g_q$ and will test the scaling theory in the limit of small $q$.

Figure \ref{fig_beta} shows $\beta(\ln g_q)$ estimated from our numerical data with $L = k_0 R$ for four different $q$. We approximate the derivative in Eq.\ (\ref{beta}) by a finite difference
\begin{eqnarray}
\beta(\ln g_q) = \frac{\partial \ln g_q(L)}{\partial \ln L} \simeq \frac{\ln g_q(L_2) - \ln g_q(L_1)}{\ln L_2 - \ln L_1}
\label{beta1}
\end{eqnarray}
and fit the result by a second order polynomial
\begin{eqnarray}
\beta(x) = A(x-x_c) + B(x-x_c)^2,
\label{poly}
\end{eqnarray}
where $x = \ln g_q$, $x_c$ is the critical value of $x$ at which $\beta = 0$, and $A$ and $B$ are constants. The fits are shown by solid lines in Fig.\ \ref{fig_beta}. The numerical results exhibit large fluctuations around the fits due to the limited number of independent disorder realizations in our data and the finite-difference approximation (\ref{beta1}) that amplifies statistical fluctuations present in the numerical data for $\ln g_q$. However, no significant systematic deviations with system size can be identified suggesting that the behavior of small-$q$ percentiles is compatible with the single-parameter scaling hypothesis of Eq.\ (\ref{beta}).

It can be shown \cite{abrahams79} that the angle at which $\beta(x)$ crosses the horizontal axis $\beta = 0$ is related to the critical exponent $\nu$ of the localization transition. More precisely, $\nu = 1/A$. Values of the critical exponent estimated in this way are given in Fig.\ \ref{fig_beta}. We see that $\nu$ decreases with $q$ in a systematic way. Although the range of $\nu = 1.52$--1.72 is compatible with $\nu \simeq 1.6$ expected for the Anderson transition in the 3D orthogonal universality class \cite{slevin14,lemarie09}, the trustworthiness of such an estimate is not satisfactory and more robust methods have to be used.

\section{Finite-size scaling of percentiles}
\label{fss}

\begin{figure*}
\includegraphics[width=0.95\textwidth]{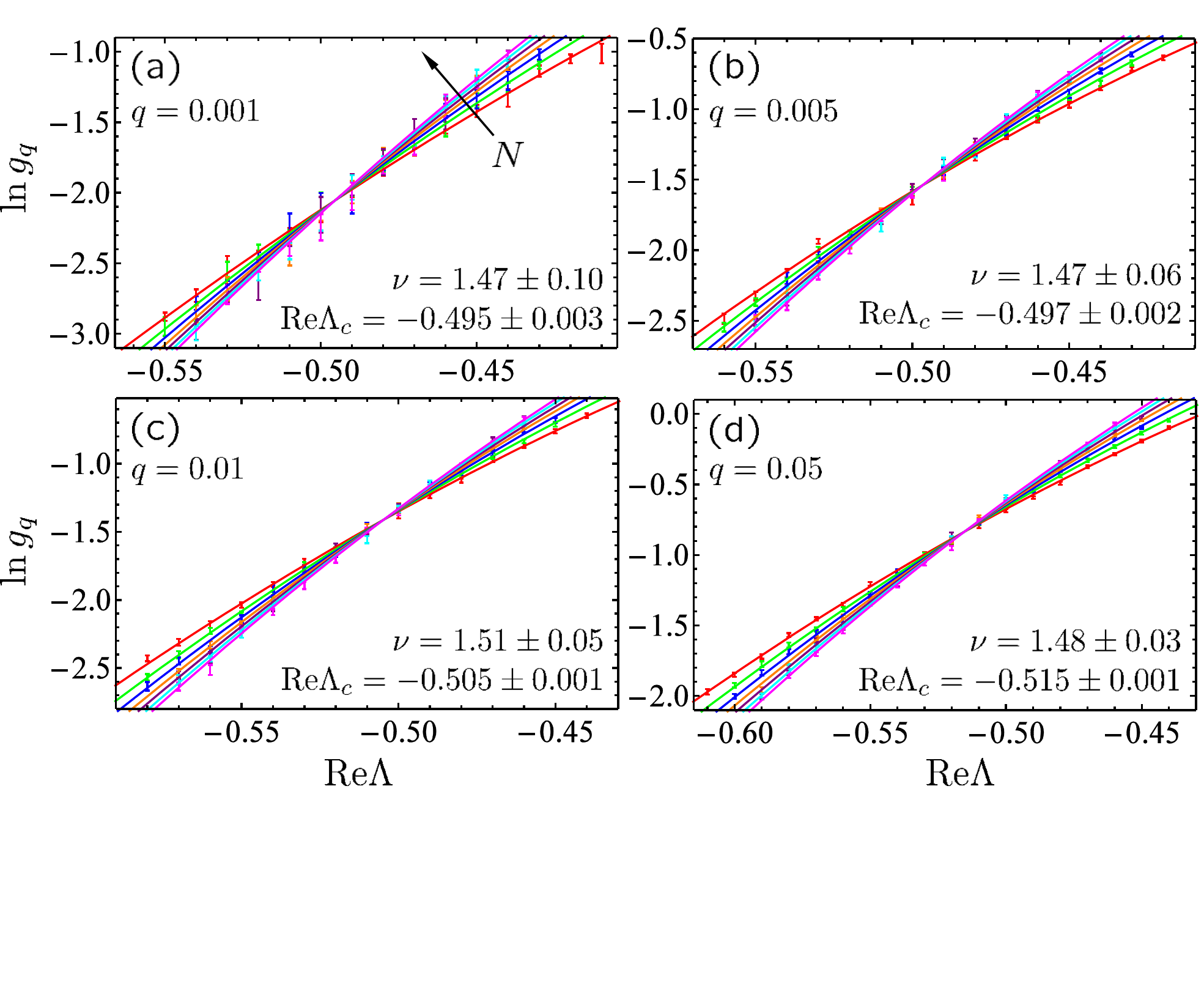}
\vspace*{-3cm}
\caption{Examples of fits (solid lines) to the numerical data (symbols with error bars) for four different values of $q$ and $m_1 = 2$, $n_1 = 1$, $m_2 = n_2 = 0$. Different curves correspond to different $N = 4$, 6, 8, 10, 12 and $16 \times 10^3$. The order of curves corresponding to increasing $N$ is indicated by an arrow in the panel (a). The best-fit parameters $\mathrm{Re}\Lambda_c$ and $\nu$ are given on the plots.}
\label{fig_fits}
\end{figure*}

The analysis presented in the previous section can be seen as a variant of finite-size scaling approach. The latter is a standard tool of statistical physics that allows estimating critical parameters of a phase transition---a phenomenon that, strictly speaking, is characteristic only for unbounded systems---from an analysis of the evolution of system's behavior with its size $L$, for finite $L$. Equation (\ref{beta}) is an appealing way to perform such an analysis because the sign of the $\beta$-function directly indicates whether the conductance grows or decreases with $L$. Unfortunately, calculating a derivative of fluctuating numerical data as suggested by this equation produces very noisy results as can be seen from Fig.\ \ref{fig_beta}. A way to perform finite-size scaling that, on the one hand, is less sensitive to statistical fluctuations of numerical data and, on the other hand, more general, was adapted to Anderson transition by Pichard and Sarma \cite{pichard81}, MacKinnon and Kramer \cite{mackinnon81}, and later perfectionized by Slevin and Ohtsuki \cite{slevin97} (see Ref.\ \onlinecite{slevin14} for a concise summary). Very generally, one hypothesizes that a quantity $\Psi$ (that has to be chosen in a ``proper'' way) obeys a single-parameter scaling. This hypothesis cannot be justified \textit{a priori} and literally means that for a large enough sample and in the vicinity of the localization transition, $\Psi$ can be parametrized by a single parameter $\phi_1$. To allow for (weak) deviations from this scaling law due to the finite sample size $L$, one introduces an additional, ``irrelevant'' scaling variable $\phi_2$ and assumes that $\Psi$ depends on the disorder strength $W$ through a scaling law of the form \cite{foot4}
\begin{eqnarray}
\Psi = F(\phi_1, \phi_2),
\label{fss1}
\end{eqnarray}
where the scaling variables are
\begin{eqnarray}
\phi_i = u_i(w) L^{\alpha_i}, \; i = 1, 2.
\label{fss2}
\end{eqnarray}
Here $w = (W-W_c)/W_c$ is the reduced disorder strength and $W_c$ is the critical disorder. The critical exponent of the disorder-driven (Anderson) transition is given by $\nu = 1/\alpha_1$. The role of the second scaling variable $\phi_2$ should decrease with $L$ (that is why it is called irrelevant) and hence $\alpha_2 = y < 0$. There may be several irrelevant scaling variables, but it is essential for our analysis that only a single relevant scaling variable $\phi_1$ exists. The validity of single-parameter scaling hypothesis is not at all guaranteed and examples of its violation exist \cite{cohen88}. The applicability of single-parameter scaling may also depend on the particular choice of the quantity $\Psi$ for which analysis is performed.

To proceed further, the functions $u_i(w)$ and $F(\phi_1, \phi_2)$ are expanded in Taylor series and one assumes that the latter can be truncated at fairly low orders:
\begin{eqnarray}
u_i(w) &=& \sum\limits_{j=0}^{m_i} b_{ij} w^j,
\label{useries}
\\
F(\phi_1, \phi_2) &=& \sum\limits_{j_1=0}^{n_1} \sum\limits_{j_2=0}^{n_2} a_{j_1 j_2} \phi_1^{j_1} \phi_2^{j_2}.
\label{fseries}
\end{eqnarray}
Typically, $m_1$, $n_1$, $m_2$, $n_2 \leq 3$ is sufficient \cite{slevin14}. Because $\Psi$ should be independent of $L$ at the critical point $W = W_c$, we must have $u_1(0) = 0$ and $b_{10} = 0$. And because Eq.\ (\ref{fseries}) contains products of $b_{ij}$ and $a_{j_1 j_2}$, we can put $a_{01} = a_{10} = 1$ without loss of generality.

We now put $W = \mathrm{Re} \Lambda$, $\Psi = \ln g_q$ and fit the numerical data corresponding to different system sizes $L = k_0 R$ and different $w$. The best fit is determined \cite{foot5} by minimizing the $\chi^2$ statistic,
\begin{eqnarray}
\chi^2 = \frac{1}{N_{\mathrm{data}}} \sum\limits_{i=1}^{N_{\mathrm{data}}}
\frac{(F_i - \Psi_i)^2}{\sigma_i^2},
\label{chi2}
\end{eqnarray}
where $\sigma_i$ is the error in the estimation of $\Psi_i$ calculated following the approach of Ref.\ \onlinecite{slevin2003} and $N_{\mathrm{data}}$ is the total number of data points. We fit only the data in an interval $\ln g_q = (\ln \tilde{g}_q)_c \pm \delta_{\ln g}$, where \{$\mathrm{Re} \tilde{\Lambda}_c$, $(\ln \tilde{g}_q)_c$\} is the approximate crossing point of curves $\ln g_q(\mathrm{Re} \Lambda, k_0R)$ corresponding to different $k_0R$ (or, equivalently, to different $N$).
$\mathrm{Re} \tilde{\Lambda}_c$ is determined as the value of $\mathrm{Re} \Lambda$ for which the sum of squared differences between $\ln g_q$ corresponding to all available $N = 2000$--16000 is minimal; $(\ln \tilde{g}_q)_c$ is the arithmetic average of $\ln g_q$ at this point.
Fits are performed for each combination of $m_1$, $n_1 \leq 3$ and $m_2 \leq m_1$, $n_2 \leq n_1$ starting with random initial values of fit parameters in ranges $\nu \in [0, 10]$, $y \in [-10, 0]$, $\mathrm{Re} \Lambda_c = \mathrm{Re} \tilde{\Lambda}_c \pm 10\%$, $a_{00} = (\ln \tilde{g}_q)_c \pm 10\%$ and all other fit parameters in a wide range $[-10, 10]$. The fitting procedure is repeated for $10^3$ different random choices of starting values and then the fit with the smallest $\chi^2$ is chosen among the fits for which the best-fit $y$ is negative and the maximal contribution of the irrelevant scaling variable does not exceed 10\%.
Fits are attempted for $\delta_{\ln g} = 0.5$, 1, and 1.5.

Analysis of the ensemble of fits shows that similar results are obtained for different $\delta_{\ln g}$, but $\delta_{\ln g} = 0.5$ implies using a very small part of available data resulting in large uncertainties of best fit parameters.  We thus choose $\delta_{\ln g} = 1$ which, on the one hand, is small enough to ensure that only the vicinity of the critical point is analyzed and, on the other hand, is sufficiently large for the part of our data falling in the range $\ln g_q = (\ln {\tilde g}_q)_c \pm \delta_{\ln g}$ to be statistically meaningful. When plotted as a function of $\mathrm{Re}\Lambda$, $\ln g_q$ for $N=2000$ significantly deviates from the curves corresponding to larger $N$ [a similar tendency for $\langle \ln g \rangle$ is seen in Fig.\ \ref{fig_lng_average}(b)] which is likely to be due to the fact that this $N$ is too small. We thus choose to discard the data corresponding to $N = 2000$ and to analyze only the data for $N \geq 4000$. To choose the orders $m_i$, $n_i$ of Taylor expansions in Eqs.\ (\ref{useries}) and (\ref{fseries}), we notice that in the majority of cases, $\chi^2$ decreases abruptly by roughly a factor of 2 or even more (and, typically, down to $\chi^2 \sim 1$) when $m_1$ or $n_1$ exceeds 2. Further increase of $m_1$, $n_1$ or introduction of $m_2$, $n_2 > 0$ does not improve fit quality significantly. Moreover, the best-fit values of the irrelevant scaling exponent $y$ obtained for $m_2$, $n_2 > 0$ vary in a very wide range for different $q$ and often become unrealistically large in absolute value, shedding doubts on the relevance of the whole analysis. We thus use $m_1 = 2$, $n_1 = 1$ and $m_2 = n_2 = 0$ (i.e., no irrelevant variable) in the remainder of this paper. The fact that we do not need the irrelevant variable to fit our data does not necessarily imply that our simulated samples are sufficiently large but, most likely, simply reflects the fact that there are still considerable statistical fluctuations in our data that exceed the eventual corrections to scaling due to the irrelevant variable.

\begin{figure}
\includegraphics[width=0.95\columnwidth]{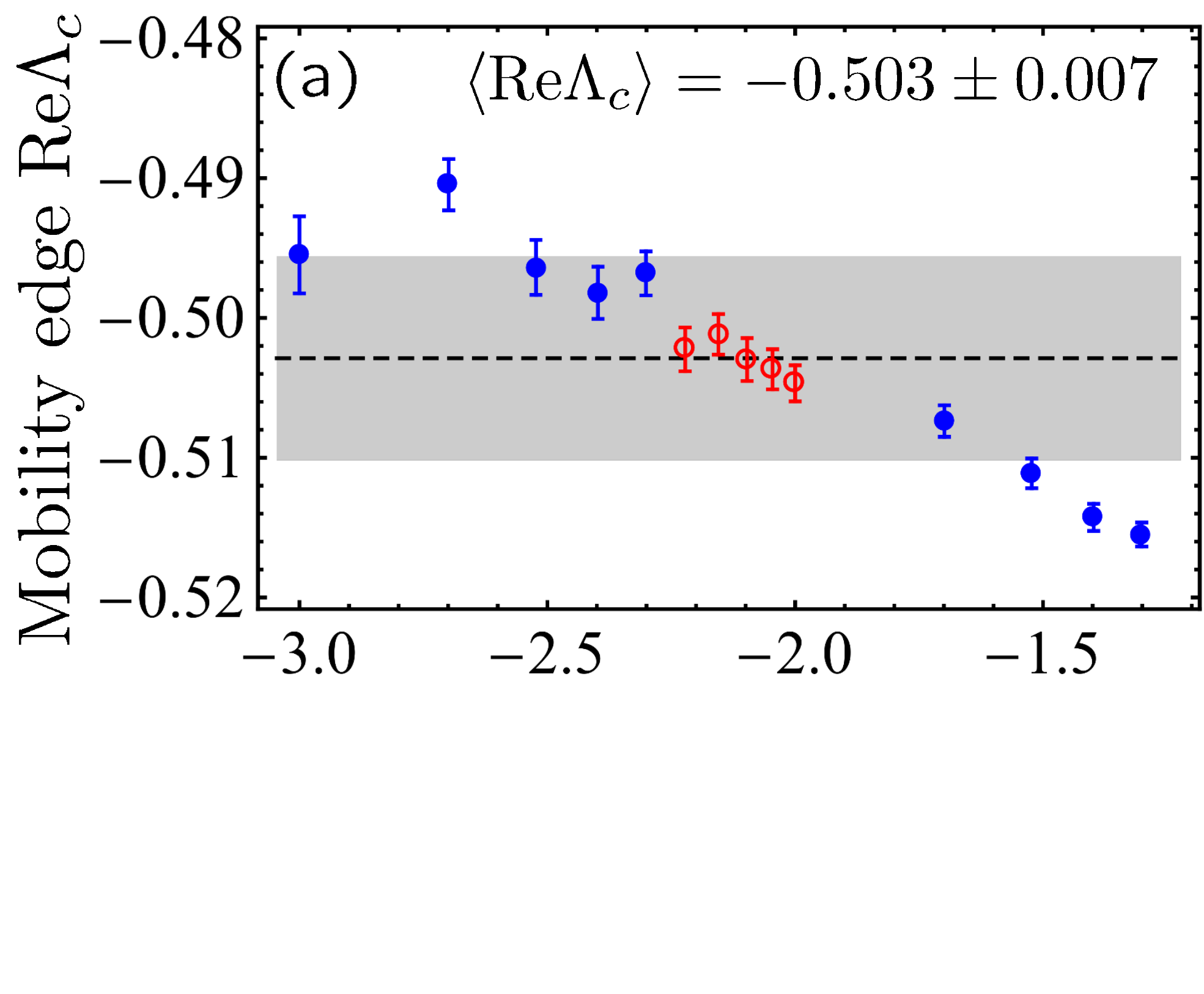}\\
\vspace*{-2.1cm}
\hspace*{1mm}\includegraphics[width=0.95\columnwidth]{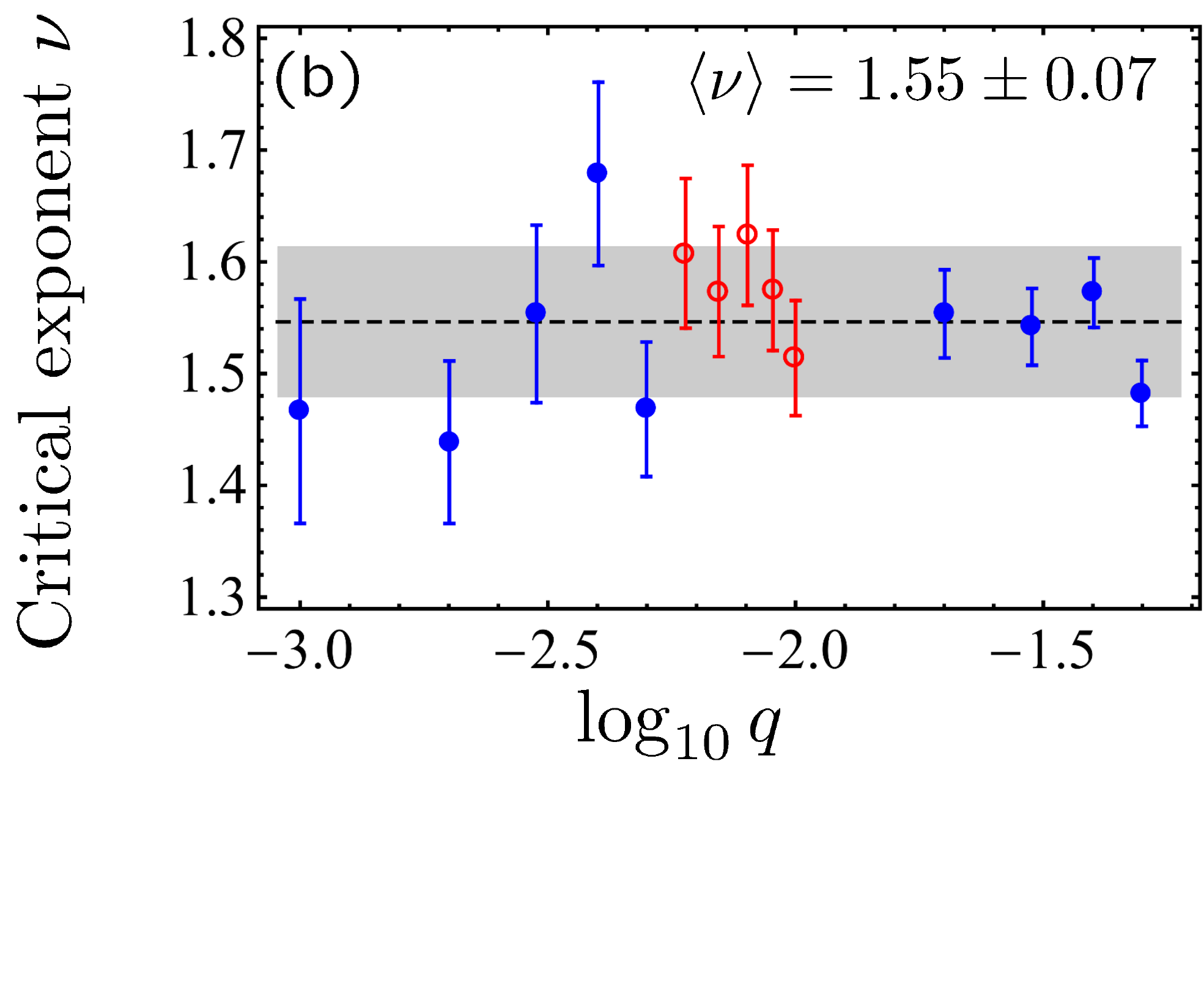}
\vspace*{-1.5cm}
\caption{Best-fit values of the mobility edge $\mathrm{Re}\Lambda_c$ (a) and of the critical exponent $\nu$ (b) obtained from the fits to the data for $q = 0.001$--$0.05$ using $m_1 = 2$, $n_1 = 1$ and $m_2 = n_2 = 0$ (symbols with error bars). The horizontal dashed lines show mean values of $\mathrm{Re}\Lambda_c$ and $\nu$, respectively. The shaded regions correspond to the uncertainties of the means.}
\label{fig_best}
\end{figure}

Examples of fits of Eqs.\ (\ref{useries}) and (\ref{fseries}) to our numerical data are shown in Fig.\ \ref{fig_fits}. The best-fit values of the mobility edge $\mathrm{Re} \Lambda_c$ and of the critical exponent $\nu$ are shown on the plots. We notice, in particular, that the estimation of the position of the mobility edge following from our analysis is significantly more precise than the estimation of the critical exponent, which is a known particularity of the finite-size scaling approach. A summary of results for different $q$ is presented in Fig.\ \ref{fig_best}. An estimate of the mobility edge following from averaging over $q = 0.001$--0.05 is $\langle \mathrm{Re} \Lambda_c \rangle = -0.503 \pm 0.007$. The average value of the critical exponent $\langle \nu \rangle = 1.55 \pm 0.07$ is consistent with $\nu \simeq 1.57$ found for the Anderson model in the 3D orthogonal universality class using the transfer matrix method \cite{slevin14} as well as with $\nu \simeq 1.59$ found for the quantum kicked rotor model that can be proved to fall in the same universality class \cite{lemarie09}. This indicates that the localization transition in the model of resonant point scatterers considered in the present work is likely to belong to the 3D orthogonal universality class as well.

Analysis of Fig.\ \ref{fig_best} suggests that there is a region of $q$, namely, $q = 0.006$--0.01, in which the values of fit parameters $\mathrm{Re} \Lambda_c$ and $\nu$ are more consistent with each other than in the full range $q = 0.001$--0.05 (points shown by open circles in Fig.\ \ref{fig_best}). This region of $q$ is likely to be the one in which our analysis is the most adequate because it corresponds to $q$ which are sufficiently small but still large enough to ensure sufficient statistical accuracy of our results. Averaging $\mathrm{Re} \Lambda_c$ and $\nu$ over this region yields better estimates of critical parameters: $\langle \mathrm{Re} \Lambda_c \rangle = -0.503 \pm 0.001$ and $\langle \nu \rangle = 1.58 \pm 0.03$. The latter value is in even better agreement with $\nu$ expected for the Anderson transition in the 3D orthogonal universality class \cite{slevin14,lemarie09}.

\begin{figure*}
\includegraphics[width=0.95\textwidth]{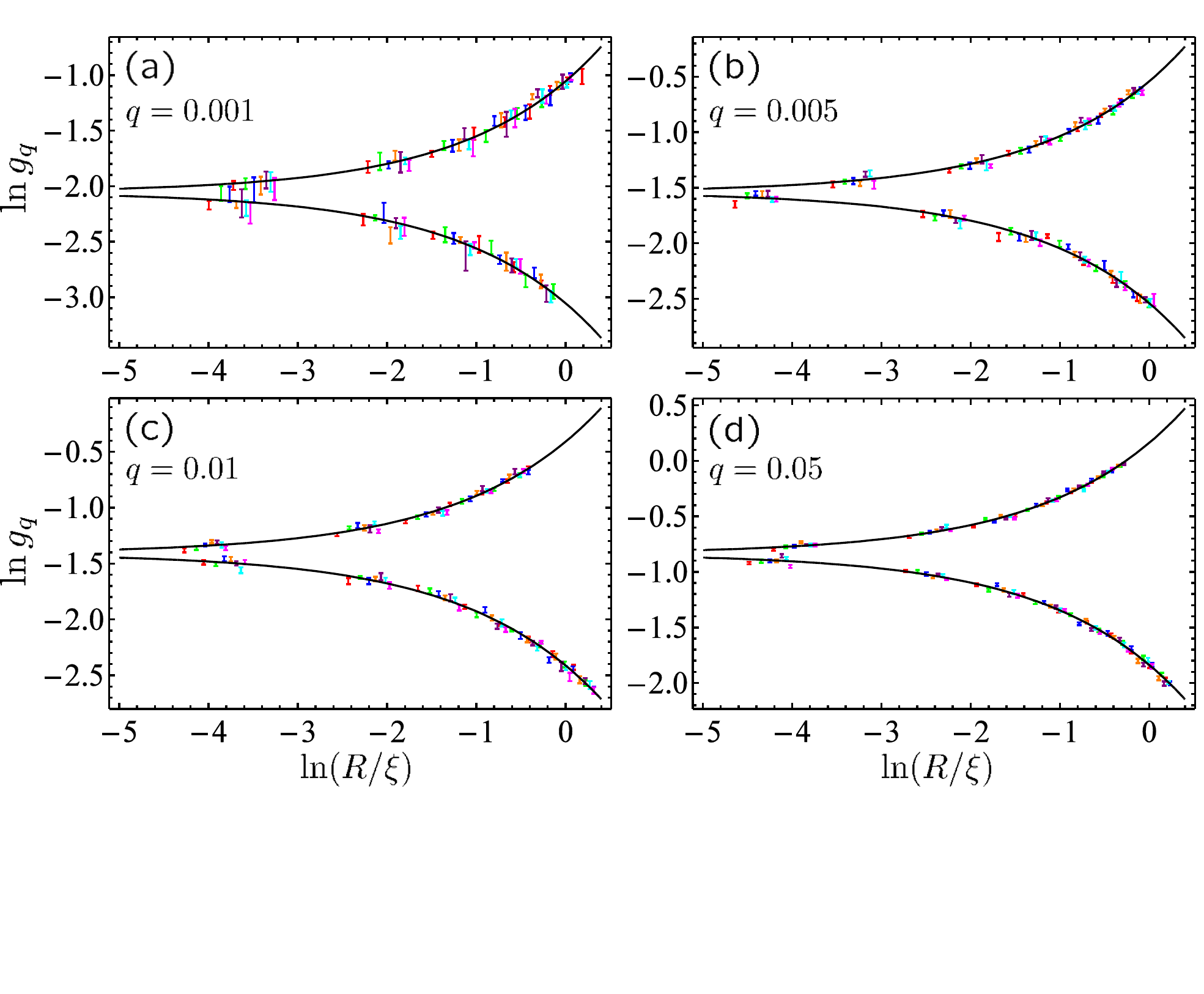}
\vspace*{-3cm}
\caption{Demonstration of single-parameter scaling of percentiles $g_q$ for four different values of $q$. Points with error bars are numerical data of Fig.\ \ref{fig_fits}. Lines show the scaling function (\ref{fss1}) as determined from the fits of Fig.\ \ref{fig_fits}.}
\label{fig_sps}
\end{figure*}

We are now in a position to check the single parameter scaling of percentiles that was one of the main assumptions of the finite-size scaling analysis presented above. For this purpose, we introduce the localization length
\begin{eqnarray}
\xi  = \frac{\mathrm{const}_{\pm}}{\left| u_1(w) \right|^{\nu}}
\label{loclength}
\end{eqnarray}
that diverges at the mobility edge where $w = 0$ and $u_1(w) = 0$. Although we know the divergence of the localization length, we cannot determine its value (set by the constants $\mathrm{const}_{\pm}$ that may be different on the two sides of the transition) from the scaling analysis performed here. We put  $\mathrm{const}_{\pm} = 1$ in the following. According to Eqs.\ (\ref{fss1}) and (\ref{fss2}), the values of $\Psi = \ln g_q$ corresponding to different sample sizes $L = k_0 R$ and disorder strengths $W = \mathrm{Re} \Lambda$ are all expected to fall on a single master curve when plotted against $L/\xi$. This is verified in Fig.\ \ref{fig_sps} where we show $\ln g_q$ as a function of $L/\xi$ for $q = 0.001$, 0.005, 0.01 and 0.05. Because numerical data nicely follow a single master curve for each $q$, we conclude that the small-$q$ percentiles $\ln g_q$ are indeed functions of a single parameter $L/\xi$ and hence our initial assumption of single-parameter scaling is verified \textit{a posteriori}.

\section{Conclusion}
\label{conclison}

We studied the disorder-induced localization transition for scalar waves in a 3D random ensemble of resonant point scatterers using the Green's matrix approach. We defined the normalized decay rates $g$ of quasi-modes of our system by dividing the imaginary part of eigenvalues $\Lambda$ of the Green's matrix ${\hat G}$ by the average spacing between $\Lambda$'s projected on the real axis. $g$ is argued to play a role analogous to Thouless conductance in electronic transport. Its statistics is analyzed with frequency resolution, i.e., by calculating statistical properties of $g$ averaged over narrow bands of $\mathrm{Re} \Lambda$. A transition between extended and localized eigenstates is observed when $\mathrm{Re} \Lambda$ crosses a critical value (mobility edge) $\mathrm{Re} \Lambda_c$, at a given and sufficiently high density of scatterers $\rho$. The probability density $p(g)$ does not obey a single-parameter scaling law and $p(g)$ obtained for different system sizes do not coincide at the critical point. We believe that this is due to other physical phenomena, such as, e.g., superradiance, that take place in our system in parallel with Anderson localization. The statistical relevance of these phenomena has a tendency to decrease with system size $L$, but we cannot claim with certainty whether they become completely irrelevant in the limit of $L \to \infty$ or not. Luckily enough, the small-$g$ parts of $p(g)$ do exhibit an $L$-independent shape for a certain value of $\mathrm{Re} \Lambda$, which suggests that small-$q$ percentiles $g_q$ of $p(g)$ may obey single-parameter scaling and hence may be used to determine the critical parameters of the localization transition. Finite-size scaling analysis of $g_q$ for $q = 0.001$--0.05 allowed us to estimate the mobility edge $\mathrm{Re} \Lambda_c = -0.503 \pm 0.007$ and the critical exponent $\nu = 1.55 \pm 0.07$ of the localization transition in our model. The latter is consistent with the value expected for the Anderson transition in the 3D orthogonal universality class. We thus conclude that the reported transition is likely to belong to this universality class as well.

The finite-size scaling analysis presented in this work can be readily extended to other wave systems described by a non-Hermitian Green's matrix model. Localization transitions in resonant scattering of elastic waves \cite{hu08,cobus15} or of light under an external magnetic field \cite{skip15} can be studied in the same way. The precise localization of one of the two mobility edges in the model (\ref{green}) achieved here opens a possibility to study the structure of critical quasi-modes by performing calculations exactly at the critical point. In particular, the multifractality of quasi-modes in open media can be investigated to improve understanding of recent experiments \cite{faez09}.

\begin{acknowledgments}
Discussions about the finite-size scaling approach with D. Delande are acknowledged. We thank the Agence Nationale de la Recherche for financial support under grant ANR-14-CE26-0032 LOVE and CNRS for support in the framework of a France-Canada PICS project Ultra-ALT.
All the computations presented in this paper were performed using the Froggy platform of the CIMENT infrastructure (https://ciment.ujf-grenoble.fr), which is supported by the Rhone-Alpes region (grant CPER07\verb!_!13 CIRA) and the Equip@Meso project (reference ANR-10-EQPX-29-01) of the programme Investissements d'Avenir supervised by the Agence Nationale de la Recherche.
\end{acknowledgments}


\begin{thebibliography}{99}
\bibitem{foot1}
Note that our definition of $g$ slightly differs from that adopted in Ref.\ \onlinecite{skip14}
\bibitem{foot3}
Note that $q = \int_0^{g_q} p_g(g) dg = \int_{-\infty}^{\ln g_q} p_{\ln g}(\ln g) d \ln g$, where $p_g(g)$ and $p_{\ln g}(\ln g)$ are probability densities of $g$ and $\ln g$, respectively.
\bibitem{foot4}
We use notations of Ref.\ \onlinecite{slevin14} except for $\Psi$ that was denoted by $\Gamma$ in Ref.\ \onlinecite{slevin14}.
\end{thebibliography}
\end{document}